\documentclass[12pt]{article}

\usepackage{amsmath,amssymb,amsfonts,color,graphicx,cite,color,soul}

\input paperdef
\newcommand{\mass}{125.5}
\newcommand{\nomix}{\emph{no-mixing}}
\newcommand{\saeff}{\emph{small $\aeff$}}
\newcommand{\gluophobic}{\emph{gluophobic Higgs}}
\newcommand{\lstop}{\emph{light stop}}
\newcommand{\lstau}{\emph{light stau}}
\newcommand{\tauphobic}{\emph{$\tau$-phobic Higgs}}
\newcommand{\lowMH}{\emph{low-$\MH$}}

\graphicspath{{figs/}}

\evensidemargin \oddsidemargin
\allowdisplaybreaks

\begin{document}

%%%%%%%%%%%%%%%%%%%%%%%%%%%%%%%%%%%%%%%%%%%%%%%%%%%%%%%%%%%%%%%%%%%%%%

\thispagestyle{empty}
\setcounter{page}{0}
\def\thefootnote{\fnsymbol{footnote}}

\begin{flushright}
ANL-HEP-PR-13-12, EFI-13-2, DESY 13-024,
FERMILAB-PUB-13-041-T\\
arXiv:1302.7033 [hep-ph]
\end{flushright}

\mbox{}\vspace{-1em}

\begin{center}

{\large\sc {\bf MSSM Higgs Boson Searches at the LHC:}}

\vspace*{0.3cm}

{\large\sc {\bf Benchmark Scenarios after the Discovery of a 
                Higgs-like Particle}}

\vspace{0.75cm}

{\sc M.~Carena$^{\,1,2}$%
%\footnote{
%email: carena@fnal.gov
%}%
, S.~Heinemeyer$^{\,3}$%
%\footnote{
%email: Sven.Heinemeyer@cern.ch
%}%
, O.~St{\aa}l$^{\,4}$%
%\footnote{
%email: oscar.stal@fysik.su.se
%}%
, C.E.M.~Wagner$^{\,2,5}$% 
%\footnote{
%email: cwagner@hep.anl.gov
%}
~and G.~Weiglein$^{\,6}$%
\footnote{
E-mail addresses: $^1$carena@fnal.gov, $^3$Sven.Heinemeyer@cern.ch, 
$^4$oscar.stal@fysik.su.se,\\ 
\mbox{}\hspace{3.3cm} $^5$cwagner@hep.anl.gov, $^6$Georg.Weiglein@desy.de
}%
}

\vspace*{0.5cm}

$^1$ Theoretical Physics Department, Fermilab,
Batavia, IL 60510-0500, USA

\vspace*{0.1cm}

$^2$ Enrico Fermi Institute and Kavli Institute for Cosmological Physics \\
Department of Physics, the University of Chicago, 5640 Ellis Ave.,
Chicago, IL 60637, USA

\vspace*{0.1cm}

$^3$ Instituto de F\'isica de Cantabria (CSIC-UC), 
     E--39005 Santander, Spain

\vspace*{0.1cm}

$^4$ The Oskar Klein Centre, Department of Physics\\ 
Stockholm University, SE-106 91 Stockholm, Sweden 
\vspace*{0.1cm}

$^5$ HEP Division, Argonne Natl.\ Lab., 9700 Cass Ave.,
Argonne, IL 60439, USA

\vspace*{0.1cm}

$^6$ DESY, Notkestra\ss{}e 85, D--22607 Hamburg, Germany

\end{center}

\vspace*{0.2cm}

\begin{abstract}
A Higgs-like particle with a mass of about $\mass \gev$ has been
discovered at the LHC. Within the current
experimental uncertainties, this new state is compatible with both the
predictions for the Standard Model (SM) Higgs boson and with the Higgs
sector in the Minimal Supersymmetric Standard Model (MSSM).
We propose new low-energy MSSM benchmark scenarios that, over a 
wide parameter range, are compatible with the mass
and production rates of the observed signal.
These scenarios also exhibit interesting phenomenology for the MSSM
Higgs sector. 
We propose a slightly updated version of the well-known 
\mhmax\ scenario, and a modified scenario (\mhmod),
where the light $\cp$-even Higgs
boson can be interpreted as the LHC signal in 
large parts of the $\MA$--$\tb$ plane. 
Furthermore, 
we define a {\it light stop scenario} that leads to a suppression
of the lightest $\cp$-even Higgs gluon fusion rate, and  a {\it light 
stau scenario} with an enhanced decay rate of $h\to \gamma\gamma$ 
at large $\tb$. We also suggest a $\tau${\it-phobic Higgs scenario}
in which the lightest Higgs can have suppressed couplings to down-type fermions.
We propose to supplement the specified value of the $\mu$~parameter in
some of these scenarios  
with additional values of both signs. This has a significant impact on
the interpretation of searches for the non SM-like MSSM Higgs bosons.  
We also discuss the sensitivity of the searches to heavy Higgs decays into
light charginos and neutralinos, and to 
decays of the form $H \to hh$. 
Finally, in addition to all the other scenarios where the lightest
$\cp$-even Higgs is interpreted as the LHC signal, we propose a
\lowMH\ scenario, where instead the {\em heavy} 
$\cp$-even Higgs boson corresponds to the new state around $\mass \gev$.
\end{abstract}

\def\thefootnote{\arabic{footnote}}
\setcounter{footnote}{0}

\newpage

%%%%%%%%%%%%%%%%%%%%%%%%%%%%%%%%%%%%%%%%%%%%%%%%%%%%%%%%%%%%%%
%%%%%%%%%%%%%%%%%%%%%%%%%%%%%%%%%%%%%%%%%%%%%%%%%%%%%%%%%%%%%%

\section{Introduction}

Elucidating the mechanism that controls electroweak symmetry
breaking (EWSB) is one of the main tasks of the LHC.
The spectacular discovery of a Higgs-like particle 
with a mass around $125$--$126 \gev$, announced
by the ATLAS and CMS experiments~\cite{ATLASdiscovery,CMSdiscovery},
marks a milestone of an effort that 
has been ongoing for almost half a century and opens a new era of
particle physics.  
Both experiments reported a clear excess in the two photon channel as
well as in the $ZZ^{(*)}$ channel, whereas 
the analyses in other channels have a lower mass resolution 
and are, at present,  less significant. The measured mass varies somewhat
between 
the different channels, and between the two experiments. We shall use the
average value $\MH^{\mathrm{obs}} = \mass \pm 1\gev$ in the following
discussion. The combined sensitivity
in each of the experiments reaches more than $ 5 \si$. The central 
value for the  observed rate in the $\ga\ga$ channel is above the expectation
for a SM Higgs boson in ATLAS results \cite{ATLASgaga}, whereas CMS measures a
lower rate \cite{CMSgaga}. Although the statistical significance of possible
deviations from the SM 
prediction is not yet sufficient to draw any definite conclusion, a confirmed
deviation in the $\ga\ga$ channel with more data could be the 
first indication of a non-SM nature of the new state, and of possible new
physics at the weak scale.

Among the most studied candidate theories for EWSB in the literature 
are the Higgs mechanism within the Standard Model (SM)~\cite{sm}
and
 the Minimal Supersymmetric Standard Model (MSSM)~\cite{mssm}. Contrary to
the SM, two Higgs doublets are required in the MSSM, resulting in five
physical Higgs boson degrees of freedom. At lowest order, where the
MSSM Higgs sector is $\cp$-conserving,
the five physical states are the light and heavy $\cp$-even Higgs
bosons, $h$ and $H$, the 
$\cp$-odd Higgs boson, $A$, and the charged Higgs boson pair, $H^\pm$.
The Higgs sector of the MSSM can be specified at lowest
order in terms of the $Z$ boson mass, $\MZ$, the $\cp$-odd Higgs mass,
$\MA$ (or the charged Higgs mass, $\MHp$), and $\tb \equiv
v_2/v_1$, the ratio of the two Higgs vacuum expectation values. 
The masses of the $\cp$-even neutral Higgs bosons and the
charged Higgs boson can be calculated,
including higher-order corrections, in terms of the other MSSM
parameters~\cite{MHreviews,PomssmRep}. An upper bound for the
mass of the lightest MSSM Higgs boson of $\Mh\lesssim 135 \gev$  was
obtained~\cite{mhiggsAEC}, and the remaining theoretical uncertainty in the
calculation of $\Mh$, from unknown higher-order corrections, was estimated 
to be up to $3 \gev$, depending on the parameter region.

Given that the experimental
uncertainties on the measurements of the production cross sections times branching ratios are still rather large, sizable deviations of various couplings from the SM values are still possible, and
even a Higgs sector that differs very significantly from the SM case can
fit the data. In particular, while within the MSSM an obvious
possibility is to interpret the new state at about $\mass \gev$ as the 
light $\cp$-even Higgs 
boson~\cite{Mh125,LightStau1,NMSSMLoopProcs,Mh125more,ADNM,Mh125evenmore}, 
it was pointed out that at least 
in principle also a much more exotic interpretation could be possible
(within the uncertainties), 
namely in terms of the {\em heavy} $\cp$-even Higgs 
boson of the MSSM~\cite{Mh125,NMSSMLoopProcs,MH125other}. 
In such a case all
five Higgs bosons of the MSSM Higgs sector would be light, where the 
heavy $\cp$-even Higgs boson would have a mass around $\mass \gev$ and 
behave roughly SM-like, while the light $\cp$-even Higgs boson of the
MSSM would have heavily suppressed couplings to gauge bosons and a mass
that would be typically below the LEP limit for a SM-like
Higgs~\cite{LEPHiggsSM}.

In parallel with the 
exciting discovery, the search for
non-standard MSSM Higgs bosons at the LHC has continued. 
The search for the
remaining Higgs bosons is pursued mainly 
via the channels ($\phi = h,H, A$):
\BEA
\label{phi-tautau}
       && p p \to \phi \to \tau^+\tau^-  
        ~(\mbox{inclusive}) , \quad 
       b \bar b \phi,  \phi \to \tau^+\tau^- 
   ~(\mbox{with}~b\mbox{-tag}) ,\\[.3em]
\label{bb-phi}
    && b \bar b \phi,  \phi \to b \bar b 
   ~(\mbox{with}~b\mbox{-tag}) ,\\[.3em]
\label{lightHpm}
     &&p p \to t \bar t \to H^\pm W^\mp \, b \bar b,~
     H^{\pm} \to \tau \nu_{\tau}~, \\[.3em]
\label{heavyHpm}
     && gb \to H^-t ~~\mbox{or}~~ g \bar b \to H^+ \bar t,~
     H^\pm \to \tau \nu_\tau~.
\EEA
The non-observation of any additional
state in these production and decay modes puts by now stringent
constraints on the MSSM parameter space, in particular on the values of
the tree-level parameters $\MA$ (or $\MHp$) and $\tb$. 
Similarly, the non-observation of supersymmetric (SUSY) particles
puts relevant constraints on the 
masses of the first and second generation scalar
quarks and the gluino, and to lesser degree on the stop and sbottom
masses (see \citere{HCP2012} for a recent summary). 

Due to the large number of 
free parameters, a complete scan of the MSSM parameter space is 
impractical in experimental analyses and phenomenological studies.
Therefore  the Higgs search results at LEP 
were interpreted~\cite{LEPHiggsMSSM} in 
several benchmark scenarios~\cite{benchmark,benchmark2}. 
In these scenarios only the two parameters that enter the Higgs sector 
tree-level predictions, $\MA$ and $\tb$, are varied (and the results
are usually displayed
in the $\MA$--$\tb$ plane), whereas the other SUSY parameters,
entering via radiative corrections, are fixed to particular benchmark
values which are chosen to exhibit certain features of the MSSM Higgs
phenomenology.
In particular, in the \mhmax~scenario the benchmark values have been
chosen such that the mass of the light $\cp$-even Higgs boson is
maximized for fixed $\tb$ and large $\MA$ (the scale of the soft
SUSY-breaking masses in the stop and sbottom sectors, which 
sets the mass scale for the corresponding supersymmetric particles, has
been fixed to $1 \tev$ in this scenario). This scenario is useful to
obtain conservative bounds on
$\tb$ for fixed values of the top-quark mass~\cite{tbexcl}.
Besides the
\mhmax\ scenario and the \nomix\ scenario, where a vanishing mixing
in the stop sector is assumed, the \saeff\ scenario and a
\gluophobic\ scenario 
were investigated~\cite{LEPHiggsMSSM}. While the latter exhibits a
strong suppression 
of the $ggh$ coupling over large parts of the $\MA$--$\tb$ parameter
space, the small~$\aeff$~scenario has strongly reduced couplings of
the light $\cp$-even Higgs boson to down-type fermions for
$\MA \lsim 350 \gev$.
This set of benchmark scenarios~\cite{benchmark,benchmark2}, 
which was originally proposed in
view of the phenomenology of the light $\cp$-even Higgs boson, was 
subsequently used also for analyses at the Tevatron and at the LHC 
in the search for the heavier MSSM Higgs bosons.  
Once the radiative corrections to the bottom mass, commonly denoted by $\De_b$, are included (see below) the predictions
for the channels used for the heavy Higgs searches are affected by a 
relevant dependence on the higgsino mass parameter $\mu$.
Hence, it 
was proposed to augment the original benchmark values of the
\mhmax\ and \nomix\ scenarios
with a variation of~$\mu$ over several discrete values (involving both
signs of $\mu$)~\cite{benchmark3}. 

The existing benchmark scenarios have provided a useful framework for
presenting limits from MSSM Higgs searches at LEP, the Tevatron
and the LHC, but those benchmark scenarios do not necessarily
permit an interpretation of 
the observed signal of a Higgs-like state at $\sim \mass \gev$ as one
of the (neutral) Higgs bosons of the MSSM Higgs sector. In particular,
the $\mhmax$ scenario has been designed such that the higher-order
corrections maximize the value of $\Mh$. As a consequence, over large parts of its
parameter space this scenario yields values of the light $\cp$-even
Higgs boson mass {\em above} the observed mass of the signal of about 
$\mass \gev$. On the other hand, the \nomix\ scenario yields 
$\Mh \lsim 122 \gev$, so that this scenario does not permit the
interpretation of the observed signal in terms of the light $\cp$-even
Higgs boson of the MSSM.
Also the other two scenarios, \saeff\
and the \gluophobic, turn out to be incompatible with $\Mh \sim \mass \gev$. 

In the present paper we therefore propose an update of the MSSM Higgs
benchmark scenarios in which we 
adapt them to the present experimental knowledge
and ongoing searches. The scenarios that we are going to propose are
defined such that over large parts of their available parameter space 
the observed signal at about $\mass \gev$ can be interpreted in terms of
one of the (neutral) Higgs bosons, while the
scenarios exhibit interesting phenomenology for the MSSM Higgs sector.

The benchmark scenarios are all specified using low-energy
MSSM parameters; we do not assume any particular soft
supersymmetry-breaking scenario. We take into account in detail the constraints from direct searches for Higgs bosons, and we
select parameters which lead to consistency with the current bounds on direct searches for
supersymmetric particles. Indirect constraints
from requiring the correct cold dark matter density, 
$\br(b \to s \ga)$, $\br(B_s \to \mu^+\mu^-$) or $(g - 2)_\mu$, however
interesting, depend to a large extent on other parameters of the theory 
that are not crucial for Higgs phenomenology. Following the spirit of
the previous benchmark proposals of 
\citeres{benchmark,benchmark2,benchmark3} we therefore do not impose
any additional constraints of this kind.
The scenarios below are defined for the MSSM with real parameters.
While an extension to complex parameters and their respective impact on the
phenomenology is interesting, it is beyond the scope of the
present paper.

The paper is organized as follows: Section~2 gives
a summary of the properties of the MSSM Higgs sector and their 
dependence on the supersymmetric parameters. In particular,
we review briefly the most important radiative corrections to the
relevant Higgs boson 
production cross sections and decay widths.
In section 3 we propose new MSSM benchmark scenarios, which update
and extend the previous benchmark proposals. We discuss the most relevant features of 
 current constraints from the LHC searches for SM-like and
non-standard Higgs bosons for each benchmark scenario, including the discovery of a Higgs-like
particle with a mass around $\mass \gev$.  
The conclusions are presented in section~4.

%%%%%%%%%%%%%%%%%%%%%%%%%%%%%%%%%%%%%%%%%%%%%%%%%%%%%%%%%%%%%%%%%%%%%%%%
%%%%%%%%%%%%%%%%%%%%%%%%%%%%%%%%%%%%%%%%%%%%%%%%%%%%%%%%%%%%%%%%%%%%%%%%

\section{Theoretical basis}

\subsection{Notation}

In the description of our notation we are including the complex
phases of the relevant SUSY parameters. However, as indicated
above, for the definition of the benchmark scenarios we restrict
ourselves to the $\cp$-conserving MSSM, i.e.\ to the case of real
parameters.
The tree-level masses of the $\cp$-even MSSM Higgs
bosons, $\Mh^{{\rm tree}}$ and $\MH^{{\rm tree}}$, 
are determined by $\tb$, the $\cp$-odd
Higgs boson mass, $\MA$, and the $Z$ boson mass, $\MZ$. The mass of the
charged Higgs boson, $\MHp^{{\rm tree}}$, 
is determined from $\MA$ and the $W$ boson
mass, $\MW$, by the relation ${(\MHp^{\rm tree})}^2=\MA^2+\MW^2$. 
The main radiative correction to the Higgs boson masses
arise from the $t/\Stop$ sector, and for large values of $\tb$ also
from the  $b/\Sbot$ and {$\tau /  \tilde{\tau}$} sectors, see
\citeres{MHreviews,PomssmRep} for reviews. 

The mass matrices for the stop and sbottom sectors of the MSSM, in
the basis of the current eigenstates $\StopL, 
\StopR$ and $\SbotL, \SbotR$, are given by
\BEA
\label{stopmassmatrix}
{\cal M}^2_{\Stop} &=&
  \ML \MstL^2 + \mt^2 + \CZb (\edz - \frac{2}{3} \sw^2) \MZ^2 &
      \mt \Xt^{*} \\
      \mt \Xt &
      \MstR^2 + \mt^2 + \frac{2}{3} \CZb \sw^2 \MZ^2 
  \MR, \\
&& \non \\
\label{sbotmassmatrix}
{\cal M}^2_{\Sbot} &=&
  \ML \MsbL^2 + \mb^2 + \CZb (-\edz + \frac{1}{3} \sw^2) \MZ^2 &
      \mb \Xb^{*} \\
      \mb \Xb &
      \MsbR^2 + \mb^2 - \frac{1}{3} \CZb \sw^2 \MZ^2 
  \MR,
\EEA
where 
\BE
\mt \Xt = \mt (\At - \mu^{*} \CTb) , \quad
\mb\, \Xb = \mb\, (\Ab - \mu^{*} \Tb) .
\label{eq:mtlr}
\end{equation}
Here $\At$ denotes the trilinear Higgs--stop coupling, $\Ab$ denotes the
Higgs--sbottom coupling, and $\mu$ is the higgsino mass parameter. We
furthermore use the notation $\sw = \sqrt{1 - \cw^2}$, with 
$\cw = \MW/\MZ$. 

SU(2) gauge invariance leads to the relation
\BE
\MstL = \MsbL .
\end{equation}
We shall concentrate on the case
\BE
\MstL = \MsbL = \MstR = \MsbR =: \msusy .
\label{eq:msusy}
\end{equation}
This identification of the diagonal elements of the third generation
squark mass matrices leads to a simple phenomenological characterization
of the third generation squark effects.  
The relaxation of this condition to the case where
$\MstR \neq \MstL \neq \MsbR$, has been studied, for instance, in
\citere{stefanCM,mhiggslong,mhiggsJRE}. 
In the case of \refeq{eq:msusy}, the most important parameters
for the corrections in the Higgs sector are  $\mt$, $\msusy$, $\Xt$, and $\Xb$. 

Similarly, the corresponding soft SUSY-breaking parameters in the scalar
tau/neutrino sector are denoted as $\Atau$ and $\msld$, where we assume
the diagonal soft SUSY-breaking entries in the stau/sneutrino mass matrices to be equal
to each other
as we did in the $\Stop/\Sbot$ sector. 
For the squarks and sleptons of the first and second generations we also
assume equality of the diagonal soft SUSY-breaking parameters,
denoted as $\msqez$ and $\mslez$, respectively. The off-diagonal
$A$-terms always appear multiplied with the corresponding fermion mass. 
Hence, for the definition of the benchmark scenarios the $A$-terms
associated with the first and second
sfermion generations have a negligible impact and can be set to zero 
for simplicity.

The Higgs sector depends also on the gaugino masses. For instance, 
at the two-loop level the gluino mass, $\mgl$, enters the
predictions for the Higgs boson masses.
The Higgs sector observables
furthermore depend on the SU(2) and U(1) gaugino mass
parameters, $M_2$ and $M_1$, respectively, which are usually
assumed to be related via the GUT relation,
\BE
M_1 = \frac{5}{3} \frac{\sw^2}{\cw^2} M_2~.
\label{def:M1}
\end{equation}

%%%%%%%%%%%%%%%%%%%%%%%%%%%%%%%%%%%%%%%%%%%%%%%%%%%%%%%%%%%%%%%%%%%%%%%%%%%%%%

\subsection{Higgs mass calculations and their scheme dependence}

Corrections to the MSSM Higgs boson sector have been evaluated in
several approaches, see, e.g.\ \citere{bse}.
The remaining theoretical uncertainty on the light $\cp$-even Higgs
boson mass has been estimated to be 
$\De \Mh^{\mathrm{theory}} \lesssim 3 \gev$ depending on the
parameter region~\cite{mhiggsAEC,PomssmRep}. 
The leading and subleading parts of the existing two-loop
calculations have been implemented into public 
codes. The program 
{\tt FeynHiggs}~\cite{feynhiggs,mhiggslong,mhiggsAEC,mhcMSSMlong} 
is based on results
obtained in the Feynman-diagrammatic (FD) approach, while the code 
{\tt CPsuperH}~\cite{cpsh} is based on results obtained using the 
renormalization group (RG)
improved effective potential
approach~\cite{mhiggsRG1a,mhiggsRG1,bse}. 
For the MSSM with real parameters the two codes can differ by a few GeV for the prediction of $\Mh$, partly due to
formally subleading two-loop corrections that are included only in 
{\tt FeynHiggs}. 
Both codes do not incorporate
the subleading two-loop contributions evaluated
in \citere{mhiggsEP5}, which are not available in a readily usable
code format. The existing 3-loop
corrections evaluated in \citeres{mh3LMartin,mh3LRobi} are also
not included, since they 
are not available in a format that can be added straight-forwardly to the
existing calculations (see, however, \citere{H3m}).

\medskip
It is important to stress that
the FD results have been obtained
in the on-shell (OS) renormalization scheme, whereas the RG results have been 
calculated using the \msbar\ scheme; a
detailed comparison of the results in the two schemes is presented in
\citeres{bse,mhiggslle} (see also \citeres{mhiggsAWN,hdecRWW}). 
Therefore, the
parameters $\Xt$ and $\msusy$ (which are most important for the
corrections in the Higgs sector) are scheme-dependent and thus 
differ in the two approaches. The
differences between the corresponding parameters
have to be taken into account  when comparing the results. Considering
the dominant standard QCD and SUSY-QCD 
corrections at the one-loop level, the relations between the stop mass
parameters in the two different schemes are given by~\cite{bse}
\BEA
\ms^{2, \MSbar} &\approx& \ms^{2, \OS} 
 - \frac{8}{3} \frac{\al_s}{\pi} \ms^2 , 
\label{eq:msms} \\
\Xt^{\MSbar} &\approx& X_t^{\OS} + \frac{\al_s}{3 \pi} \ms 
   \left(8 + 4 \frac{X_t}{\ms} - \frac{X_t^2}{\ms^2}
   - 3 \frac{X_t}{\ms} 
   \log\left(\frac{\mt^2}{\ms^2}\right) \right),
\label{eq:xtms} 
\EEA 
where $\ms^2 := \msusy^2 + \mt^2$. In these relations we have assumed
$\mgl=\msusy$. It should be noted that it is not necessary to
distinguish between \msbar\ and on-shell quantities in the terms
proportional to $\al_s$, since this difference is of higher order.
The change of scheme induces in general only a minor shift, of the order
of 4\%, in the parameter $\msusy$, but
sizable differences can occur between the 
numerical values of 
$\Xt$ in the two schemes, see \citeres{mhiggslong,bse,hdecRWW}.

%%%%%%%%%%%%%%%%%%%%%%%%%%%%%%%%%%%%%%%%%%%%%%%%%%%%%%%%%%%%%%%%%%%%%%%%

\subsection{Leading effects from the bottom/sbottom sector}

At tree level, the bottom quark Yukawa coupling,
$h_b$, controls the interaction between the Higgs fields and
the sbottom quarks and determines the bottom quark mass $\mb =h_b v_1$.
This relation is affected at \onel\ order by large radiative
corrections proportional to $h_b v_2$
\cite{deltamb1,deltamb2,deltamb2b,deltamb3}, thereby 
giving rise to $\tb$-enhanced contributions.
These terms, that are often called threshold
corrections to the bottom quark mass or $\De_b$ corrections, may be generated
 by gluino--sbottom \onel\ diagrams (resulting in \order{\alb\als}
corrections to the Higgs masses, where $\alpha_b=h_b^2/4\pi$),  by
chargino--stop loops (giving \order{\alb\alt} corrections, where
$\al_t=h_t^2/4\pi$), or by 
other subleading contributions.  At
sufficiently large values of $\tb$, the $\tb$-enhancement may
compensate the loop suppression, and these
contributions may be numerically relevant. Therefore, an accurate determination
of $h_b$ from the experimental value of the bottom quark mass requires a
resummation of these threshold effects to all orders in the perturbative
expansion \cite{deltamb2,deltamb2b}.

The leading $\De_b$-induced effects on the Higgs couplings may be
included in an effective Lagrangian formalism~\cite{deltamb4,deltamb2}.
Numerically this represents the dominant
contributions to the Higgs couplings from the sbottom sector (see also
\cite{mhiggsEP4,mhiggsEP4b,mhiggsFD2}). 
The effective Lagrangian is given by
\begin{align}
\label{effL}
\cL = \frac{g}{2\MW} \frac{\mbms}{1 + \db} \Bigg[ 
&\quad \tb\; A \, i \, \bar b \ga_5 b 
   + \wz \, V_{tb} \, \tb \; H^+ \bar{t}_L b_R \\
&+ \KL \frac{\Sa}{\Cb} - \db \frac{\Ca}{\Sbe} \KR h \bar{b}_L b_R 
                                                              % \non \\
%&-& 
- \KL \frac{\Ca}{\Cb} + \db \frac{\Sa}{\Sbe} \KR H \bar{b}_L b_R
    \Bigg] + \!{\rm h.c.}~. \non
\end{align}
Here $\mbms$ denotes the running bottom quark mass at the chosen scale including SM QCD
corrections. 
The prefactor $1/(1 + \db)$ in \refeq{effL} arises from the
resummation of the leading corrections to all orders. 
The additional terms proportional to $\db$ in the $h\bar b b$ and $H\bar b b$
couplings arise from the mixing between the $\cp$-even Higgs bosons and
from the one-loop coupling of the bottom quark to $H_u$ (the doublet
that gives masses to the up-type fermions).

As stressed above there are two main contributions to
the threshold correction $\db$, 
an \order{\als} correction from a
sbottom--gluino loop and an \order{\alt} correction
from a stop--higgsino loop.
In the limit of $M_S \gg \mt$ and $\tb \gg 1$, taking these two
contributions into account%
\footnote{
The evaluation in {\tt FeynHiggs} that we shall use 
in our numerical computations contains the full one-loop
contributions to $\db$ as given in~\citere{deltambfull}.  The
leading QCD two-loop corrections to $\db$ are also
available~\cite{deltamb2L}; they stabilize the scale dependence of
$\db$ substantially. Corrections in the MSSM with non-minimal
  flavor structure were recently published in \citere{dmb-nmfv}.}%
~$\db$ reads~\cite{deltamb1}
\BE
\db = \frac{2\als}{3\,\pi} \, \mgl \, \mu \, \tb \,
                    \times \, I(\msbe, \msbz, \mgl) +
      \frac{\alt}{4\,\pi} \, \At \, \mu \, \tb \,
                    \times \, I(\mste, \mstz, \mu) ~.
\label{def:dmb}
\end{equation}
The function $I$ is given by
\BEA
I(a, b, c) &=& \ed{(a^2 - b^2)(b^2 - c^2)(a^2 - c^2)} \,
               \KL a^2 b^2 \log\frac{a^2}{b^2} +
                   b^2 c^2 \log\frac{b^2}{c^2} +
                   c^2 a^2 \log\frac{c^2}{a^2} \KR \\
 &\sim& \ed{\mbox{max}(a^2, b^2, c^2)} ~. \non
\EEA

The $\db$ correction can become very important for large values of $\tb$
and the ratios of $\mu \mgl/ \msusy^2$ and  $\mu \At/ \msusy^2$.
While for $\mu, \mgl, \At > 0$,  the $\db$ correction is positive, 
leading  to a suppression of the bottom Yukawa coupling, 
for negative values of $\db$ the bottom Yukawa coupling
may be strongly enhanced and can even acquire 
non-perturbative values when $\db \to -1$.

The impact of the $\db$ corrections on the
searches for the heavy MSSM Higgs bosons has been
analyzed in \citere{benchmark3} (see also
\citeres{cmsHiggs,cmsHiggs2}). It was shown that the  
exclusion bounds in the channels defined by \refeqs{bb-phi} and
(\ref{lightHpm}) 
depend strongly on the sign and size of $\db$, whereas the channels
\refeqs{phi-tautau} and (\ref{heavyHpm}) 
show a weaker dependence on $\db$, as a consequence of a partial
cancellation of the $\db$ contributions. In order to demonstrate the
phenomenological consequences of varying the parameter $\mu$,
it was recommended in~\citere{benchmark3} 
to augment the original benchmark values of the \mhmax\ and \nomix\
scenarios~\cite{benchmark2} with a variation of $\mu$ over discrete
values in the range $-1000 \gev$ to $+1000 \gev$. When investigating
negative values of $\mu$, in particular $\mu = -1000\gev$, the
considered range of $\tb$ needs to be restricted to sufficiently low
values in order to maintain a perturbative behavior of the bottom Yukawa
coupling.

%%%%%%%%%%%%%%%%%%%%%%%%%%%%%%%%%%%%%%%%%%%%%%%%%%%%%%%%%%%%%%%%%%%%%%%%%
%%%%%%%%%%%%%%%%%%%%%%%%%%%%%%%%%%%%%%%%%%%%%%%%%%%%%%%%%%%%%%%%%%%%%%%%%

\section{Benchmark Scenarios}

In the following subsections we propose updated benchmark scenarios, in
which the observed LHC signal at $\sim \mass \gev$ can be interpreted as
one of the (neutral $\cp$-even) states of the MSSM Higgs sector, and we
discuss relevant features of their phenomenology. In particular, within
present experimental uncertainties, these benchmark scenarios allow for
different interpretations of the production and decay rates of the
discovered Higgs-like state. In addition, the scenarios are useful in
the search of the other, non SM-like, MSSM Higgs bosons. 
For convenience, we also give a table
containing the parameter values for all the proposed
scenarios in the Appendix.

Concerning the parameters that have only a minor impact on the MSSM Higgs
sector predictions, we propose fixing them to the following values:
\begin{align}
\msqez &= 1500 \gev, \\
\mslez &= 500 \gev, \\
A_f &= 0 \quad (f = c,s,u,d,\mu,e)~.
\label{LHCcolored}
\end{align}
$M_1$ is fixed via the GUT relation, \refeq{def:M1}.
Motivated by the analysis in Ref. \cite{benchmark3} we suggest to investigate
for each scenario given in \refses{sec:mhmax} --
\ref{sec:lightstop}, in addition to the default values given there, 
the following values of $\mu$:
\begin{align}
\mu = \pm 200, \pm 500, \pm 1000 \gev.
\label{eq:musuggest}
\end{align}
These values of $\mu$
allow for both an enhancement and a suppression of the bottom Yukawa
coupling, and are consistent with the limits from direct searches for
charginos and neutralinos at LEP~\cite{pdg}.
As mentioned above, when investigating
negative values of $\mu$ the
considered range of $\tb$ needs to be restricted to sufficiently low
values in order to maintain a perturbative behavior of the bottom Yukawa
coupling.

The value for the top quark mass used in the original benchmark
scenarios~\cite{benchmark2,benchmark3} was chosen according to the
experimental central value at that time. For the new scenarios we
propose to substitute this value with the most up-to-date experimental
central value $\mt=173.2\gev$ \cite{mt1732}.

\medskip
To analyze the benchmark scenarios discussed below, and to generate the
MSSM Higgs predictions for the plots, we use  
{\tt FeynHiggs\,2.9.4}~\cite{feynhiggs,mhiggslong,mhiggsAEC,mhcMSSMlong}. 
Where relevant, values for the input parameters are quoted both in the
on-shell scheme (suitable for {\tt FeynHiggs}), as well as in the
\msbar\ scheme. The latter set of parameters can readily be used by 
{\tt CPsuperH}~\cite{cpsh}. Using this code we have verified that these
parameter settings lead to similar Higgs phenomenology.%
\footnote{For calculations of the Higgs branching ratios, there also exist
  other codes like {\tt HDECAY} \cite{hdecay}. The branching ratio predictions
  for the different scenarios are generally in good agreement between the
  different codes, and we use {\tt FeynHiggs} for simplicity.}% 
~We also show the exclusion bounds (at $95\%$ C.L.) from direct Higgs searches,
evaluated with {\tt HiggsBounds\,4.0.0}~\cite{higgsbounds,HBmanual} 
(linked to {\tt FeynHiggs}). This code uses exclusion limits from LEP,
the Tevatron, and the LHC (results presented up until the Moriond 2013
conference are included). In particular this includes the most sensitive
limits from searches for neutral~\cite{HCP2012tautau,ATLASHtautau} and
charged \cite{ChargedHiggs} MSSM Higgs bosons, and the combined limits on
Higgs bosons with SM-like couplings \cite{ATLASdiscovery,CMSSMHiggs}. For a
full list of included limits and references, we refer to Appendix A of
\citere{HBmanual}. 
A combined uncertainty on the SM-like Higgs mass
of $\De \Mh = 3 \gev$ ($\De \MH = 3 \gev$ in the last scenario) was used when
evaluating the limits. While an estimate of the currently excluded
region is given in this way,%
\footnote{
{\tt HiggsBounds} provides a
compilation of cross section limits obtained from Higgs searches at LEP,
the Tevatron and the LHC. For testing whether a particular parameter
point of a considered model is excluded, first the search channel with the
highest expected sensitivity for an exclusion is determined, and then
the observed limit is confronted with the model predictions 
for this single channel only, see \citere{higgsbounds} for further
details.}
 we would like to
emphasize that a main point of this work is to encourage ATLAS and CMS
to perform dedicated searches for MSSM Higgs bosons in these scenarios. 

For each benchmark scenario we show the region of parameter
space where the mass of the (neutral $\cp$-even) MSSM Higgs boson 
that is interpreted as the newly discovered state is within the 
range $\mass \pm 3 \gev$ and $\mass \pm 2 \gev$.
The $\pm 3\gev$ uncertainty is meant to
represent a combination of the present experimental uncertainty of the
determined mass value and of the
theoretical uncertainty in the MSSM Higgs mass prediction from unknown
higher-order corrections. 
Taking into account a parametric uncertainty from the top quark mass
measurements of $\de\mt^{\rm exp} = 0.9 \gev$~\cite{mt1732} would result in
an even slightly larger interval
of ``acceptable'' $\Mh$ values, while all other features remain the same.
The displayed area with $\pm 3\gev$
uncertainty should therefore be viewed as being in (conservative)
agreement with a Higgs mass measurement of $\sim \mass \gev$. 
In particular, in the case that the lightest $\cp$-even Higgs is
interpreted as the newly discovered state, the couplings of the $h$
are close to the corresponding SM values (modulo effects from light
SUSY particles, see below). Consequently, those rate measurements from
the LHC that agree well with the SM
are then naturally in good agreement also with the MSSM predictions.
The area corresponding to the $\pm 2\gev$ uncertainty indicates how the
region that is in agreement with the measured value would shrink as a
consequence of reducing the theoretical and experimental uncertainties
to a combined value of $2\gev$.

%%%%%%%%%%%%%%%%%%%%%%%%%%%%%%%%%%%%%%%%%%%%%%%%%%%%%%%%%%%%%%%%%%%%%%%%%%%%%%%

%\newpage
\subsection{The \boldmath{$\mhmax$} scenario}
\label{sec:mhmax}

The \mhmax\ scenario was originally
defined to give conservative exclusion bounds
on $\tb$ in the LEP Higgs 
searches~\cite{benchmark2,tbexcl,LEPHiggsMSSM}. The value 
of $X_t$ was chosen in order to maximize the lightest $\cp$-even Higgs
mass at large values of $\MA$ for a given value of $\tb$ (and all
other parameters fixed). Taking into account (besides the latest limits
from the Higgs searches at the Tevatron and the LHC) 
the observation of a new state at 
$\sim \mass \gev$ and interpreting this signal
as the light $\cp$-even Higgs, the \mhmax\ scenario can now be used to
derive conservative lower bounds on $\MA$, $\MHp$ and $\tb$~\cite{Mh125}.

On the other hand, since 
the $\mhmax$ scenario has been designed such that the higher-order
corrections maximize the value of $\Mh$, in the decoupling region
($\MA \gg \MZ$) and for $\tb \gsim 10$ this scenario yields $\Mh$ values
that are significantly higher (above 130~GeV) than the observed mass of
the signal. Compatibility of the predicted values for the mass of the
light $\cp$-even Higgs boson with the mass of the observed signal is therefore achieved only in a relatively small region of the parameter space, in particular for rather low
values of $\tb$. 
However, given that the $\mhmax$ scenario is useful to provide conservative lower bounds on the
parameters determining the MSSM Higgs sector at tree level ($\MA$ or
$\MHp$ and $\tb$) and has widely been used for analyses in the past, we nevertheless regard it as a useful benchmark scenario also for the future. We
therefore include a slightly updated version of the $\mhmax$ scenario in
our list of proposed benchmarks.

We define the parameters of the (updated) \mhmax\ scenario  (with the
remaining values as defined in the previous section) as follows,

\underline{\mhmax:}\\[-2em]
\begin{align}
\mt &= 173.2 \gev, \non \\
\msusy &= 1000 \gev, \non \\
\mu &= 200 \gev, \non \\
M_2 &= 200 \gev, \non \\
\Xt^{\OS} &= 2\, \msusy  \; \mbox{(FD calculation)}, \non \\
\Xt^{\MSbar} &= \sqrt{6}\, \msusy \; \mbox{(RG calculation)}, \non \\ 
\Ab &= \Atau = \At, \non \\
\mgl &= 1500 \gev, \non \\
\msld &= 1000 \gev~.
\label{mhmax}
\end{align}
Besides (as mentioned above) using the current
experimental central value for the top quark mass, the most relevant
change in the definition of the \mhmax\ scenario is an increased value
of the gluino mass, which has been adopted in view of the limits from
the direct searches for SUSY particles at the LHC~\cite{HCP2012}.
It should be noted that 
slightly higher values of $\Mh$ can be reached if one uses lower values of
$\mgl$ as input.
Consequently, slightly {\em more conservative} exclusion bounds on $\tb$,
$\MA$ and $\MHp$ can be obtained 
if one uses as input the lowest possible value for $\mgl$ that is still allowed in this scenario by the 
most up-to-date exclusion bounds from ATLAS and CMS, but with
$\mgl \ge 800 \gev$.
Similarly, {\em more conservative} exclusion bounds can of course also
be obtained by increasing the input value for $\msusy$, for
instance by using 
$\msusy = 2000 \gev$ and $\mgl = 0.8\, \msusy$ (i.e., the
``original'' setting of $\mgl$ as defined in \citere{benchmark2}), see
below. We encourage 
the experimental collaborations to take into consideration in their
analyses also those extensions of the \mhmax\ scenario.

%%%%%%%%%%%%%%%%%%%%%%%%% F I G U R E %%%%%%%%%%%%%%%%%%%%%%%%%%%%%%%%%%%%%%%%%
\begin{figure}[tbh!]
%\vspace{2em}
\begin{center}
\includegraphics[width=0.45\textwidth]{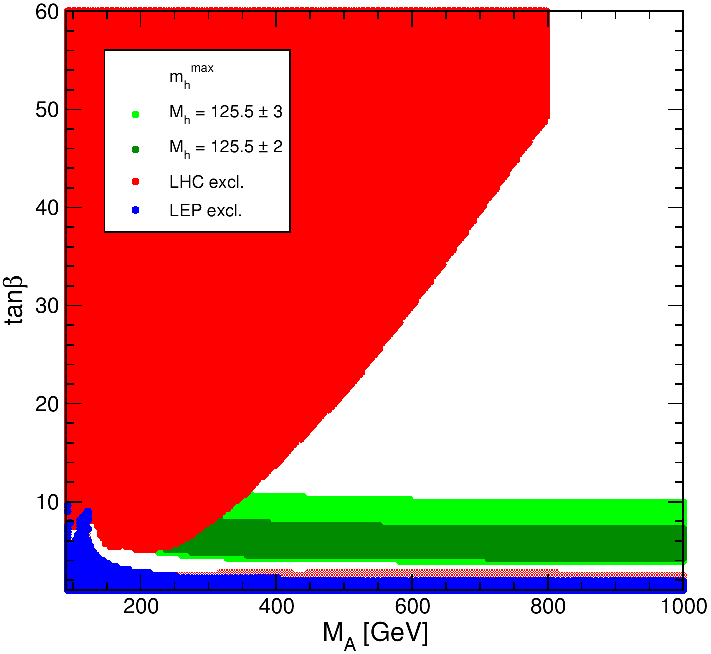}
\includegraphics[width=0.45\textwidth]{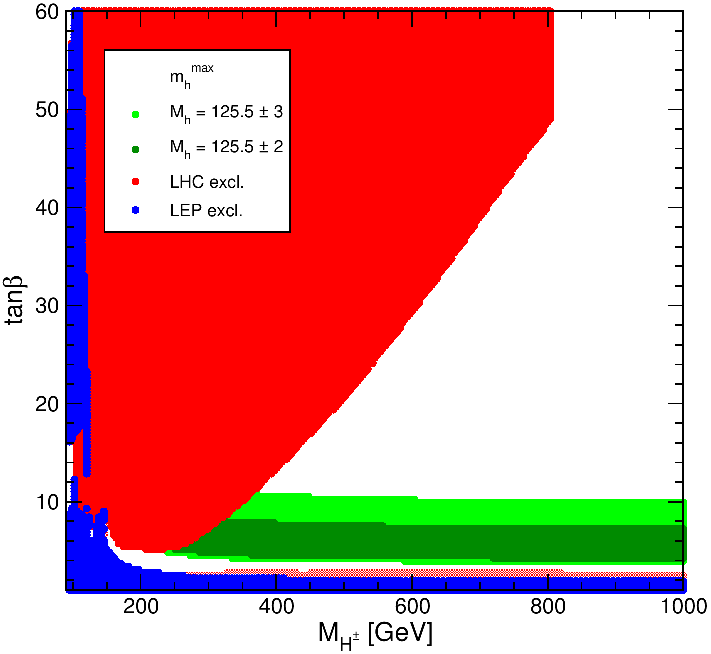}
\caption{
The $\MA$--$\tb$ (left) and $\MHp$--$\tb$ (right) planes in the (updated)
\mhmax\ scenario, with excluded regions from direct Higgs searches at
LEP (blue), and the LHC (solid red); the dotted (lighter) red 
region is excluded by
  LHC searches for a SM-like Higgs boson. The two green shades 
correspond to the parameters for which $\Mh = \mass \pm 2\, (3) \gev$,
see text.
}
\label{fig:mhmax}
\end{center}
%\vspace{-2em}
\end{figure}
%%%%%%%%%%%%%%%%%%%%%%%%% F I G U R E %%%%%%%%%%%%%%%%%%%%%%%%%%%%%%%%%%%%%%%%%

In \reffi{fig:mhmax} we show the
$\MA$--$\tb$ plane (left) and the $\MHp$--$\tb$ plane (right) 
in the (updated) \mhmax\ scenario.  
As explained above, the areas marked as excluded in the plots
have been determined using 
{\tt HiggsBounds\,4.0.0-beta}~\cite{higgsbounds} (linked to  
{\tt FeynHiggs}).
The blue areas in the figure indicate regions that
are excluded by LEP Higgs searches, and
the red areas indicate regions that are 
excluded by LHC searches for a 
SM Higgs (lighter red) and for (non-standard)
MSSM Higgs bosons (solid red). The solid red region of
LHC exclusion in this plane cuts in from
the upper left corner, in the region of large $\tb$. The most
sensitive processes here are given by
\refeq{phi-tautau}. These processes have an 
enhanced rate growing with $\tb$. The ``cutoff'' in the excluded region 
for $\MA>800\gev$ (corresponding roughly to values of $\tb$ above 50) 
is due to the fact that no experimental limits for $\MA>800\gev$ have
yet been published.

Furthermore, \reffi{fig:mhmax} shows regions in lighter red (``thin
strips'' at $\tb$ values close to the LEP limit and moderate to large values
of $\MA$ and $\MHp$), indicating
the exclusion of the light $\cp$-even Higgs boson via SM-Higgs searches at
the LHC.
In this region the LHC
extends the LEP exclusion bounds for a SM-like Higgs
to higher Higgs boson masses.

\begin{figure}[tbh!]
%\vspace{2em}
\begin{center}
\includegraphics[width=0.45\textwidth]{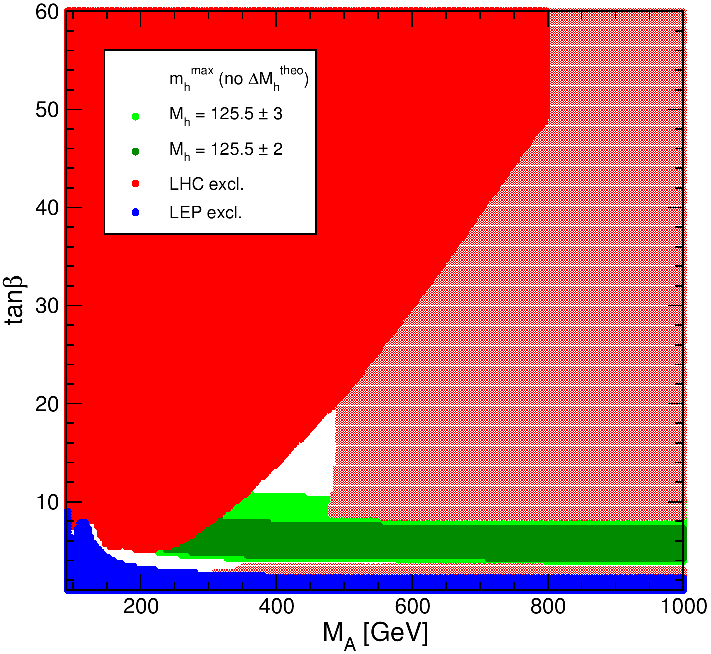}
\includegraphics[width=0.45\textwidth]{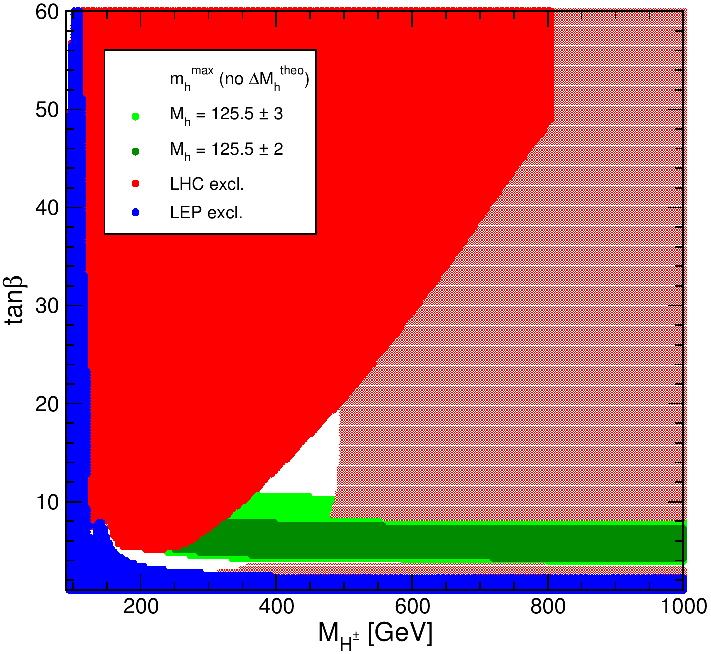}
\caption{
The $\MA$--$\tb$ (left) and $\MHp$--$\tb$ (right) planes in the (updated)
\mhmax\ scenario, as shown in \reffi{fig:mhmax} (using the same color
coding), but without taking
   into account a theory uncertainty in the $\Mh$ calculation of $3 \gev$ 
  in the evaluation of the existing limits.
}
\label{fig:mhmax-nodMh}
\end{center}
%\vspace{-2em}
\end{figure}

The two green colors in \reffi{fig:mhmax} indicate where $\Mh =
\mass \pm 2\, (3) \gev$.  
As discussed above, the $\pm 3\gev$ region should represent a
reasonable combination of the current experimental and theoretical
uncertainties. The fact that the LHC exclusion region from the SM
Higgs searches does not exactly ``touch'' the
green band is a consequence of taking into account the theoretical 
uncertainties in the prediction for the Higgs boson mass in determining
the excluded regions. The incorporation of the theoretical uncertainties
is also responsible for the fact that in \reffi{fig:mhmax} there is no 
excluded region from the SM Higgs searches at the LHC for $\tb$ values
above the green region. It may be useful to regard the green region
as that favored by the LHC observation, even though other parameter
regions exist that are not formally excluded (according to the
prescription adopted in {\tt HiggsBounds}~\cite{higgsbounds}).
The effects of the theory uncertainty of $\pm 3 \gev$ used in the
evaluation of the experimental bounds 
are displayed in \reffi{fig:mhmax-nodMh}, where we {\em neglect} this theory
uncertainty. It can be observed that large parts of the $\MA$--$\tb$
plane (left) and of the $\MHp$--$\tb$ plane (right) would then be
excluded in the \mhmax\ scenario from the LHC searches for a
SM-like Higgs boson. The resulting 
excluded region is shown in light red.
In particular, for $\tb$ values above the green band the predicted
  $\Mh$ value turns out to be {\em too high}.

%%%%%%%%%%%%%%%%%%%%%%%%% F I G U R E %%%%%%%%%%%%%%%%%%%%%%%%%%%%%%%%%%%%%%%%%
%%%%%%%%%%%%%%%%%%%%%%%%% F I G U R E %%%%%%%%%%%%%%%%%%%%%%%%%%%%%%%%%%%%%%%%%

Interpreting the light $\cp$-even Higgs as the new state at 
$\sim \mass \gev$, a new conservative lower bound on $\tb$ in the MSSM
can be obtained from the lowest values on the green bands in 
\reffi{fig:mhmax} (see \citere{Mh125} for details).
Similarly, the lowest values of $\MA$ and $\MHp$ in the green region
(i.e., where the green region touches the excluded region from Higgs
searches at the LHC)
give a conservative lower bound on these parameters~\cite{Mh125}. In
particular, from the right plot of \reffi{fig:mhmax} it follows that 
$\MHp < \mt$ is excluded for $\msusy=1\tev$ (if the light $\cp$-even Higgs is interpreted
as the new state at $\sim \mass \gev$). 
Raising $\msusy$ to higher values, e.g.\ to $2000 \gev$, one finds
that $\MHp < \mt$ might still be marginally allowed.
These bounds could be improved by a more precise theoretical
prediction and experimental determination of $\Mh$, and more data on
MSSM Higgs boson searches in the region of low values of $\MA$ could
clearly have an important impact.

It should finally be noted that the sensitivity of the 
searches for MSSM Higgs bosons in 
$\tau^+\tau^-$ and $b \bar b$ final states that determines the solid red
region in \reffi{fig:mhmax} is significantly affected where additional
decay modes of the heavy MSSM Higgs bosons are open. In particular, for
sufficiently large values of $\MA$ decays of the MSSM Higgs bosons $H$
and $A$ into charginos and neutralinos can have an important impact,
depending on the parameters in the chargino/neutralino sector. 
This issue will be discussed in more detail below. Furthermore, 
interpreting the light $\cp$-even Higgs as the new state at
$\sim \mass \gev$ means that the decay $H \to hh$ is kinematically
possible over a large part of the parameter space of the \mhmax\
scenario (and of its variants that will be discussed below). 
This decay mode can be particularly important in the region of
relatively low values of $\tb$ that is favored in the \mhmax\ scenario
(see \citeres{hhh,hdecRWW} for details of the calculation.) 
As an example, for $\MA = 300 \gev$ and $\tb = 7$, i.e.\ close to the
experimental limit from the Higgs searches at the LHC, we find 
$\br(H \to hh) = 12\%$. This branching ratio increases for
lower values of $\tb$. For $\tb = 4.5$ we find 
$\br(H \to hh) = 27\%$. The two values quoted above are for 
$M_2 = 200 \gev$, where also competing decay modes into charginos and
neutralinos are open. Increasing the SU(2) 
gaugino mass parameter to $M_2 = 2000 \gev$, thus increasing the masses
of the charginos and neutralinos, yields $\br(H \to hh) = 19\%$ for 
$\tb = 7$ and $\br(H \to hh) = 50\%$ for $\tb = 4.5$ (for $\MA = 300 \gev$, 
as before).
We encourage ATLAS and CMS to enhance the sensitivity of their searches
for MSSM Higgs bosons by performing also dedicated searches for Higgs
decays into SUSY particles (see the discussion below), where initial
  analyses can be found, e.g., in \citere{CMSchichi}.

%%%%%%%%%%%%%%%%%%%%%%%%%%%%%%%%%%%%%%%%%%%%%%%%%%%%%%%%%%%%%%%%%%%%%%%%%%%%%%%

\subsection{The \boldmath{\mhmod} scenario}
\label{sec:mhmod}

As explained in the discussion of \reffi{fig:mhmax}, the mass of 
the light $\cp$-even Higgs
boson in the \mhmax\ scenario is in 
agreement with the discovery of a Higgs-like state only in a
relatively small strip
in the $\MA$--$\tb$ plane at rather low $\tb$. 
This was caused by the fact that the $\mhmax$ scenario was designed 
to maximize the value of $\Mh$, so that in the decoupling region this
scenario yields $\Mh$ values that are higher than the observed mass of
the signal. Departing from the parameter configuration that maximizes 
$\Mh$, one naturally finds scenarios where in the decoupling region the 
value of $\Mh$ is close to the observed mass of the signal over a wide
region of the parameter space. A convenient way of modifying the 
$\mhmax$ scenario in this way is to reduce the amount of mixing in the
stop sector, i.e.\ to reduce $|\Xt/\msusy|$ compared to the value of
$\approx 2$ (FD calculation) that gives rise to the largest positive
contribution to $\Mh$ 
from the radiative corrections. This can be done for both signs of
$\Xt$. 

Accordingly, we propose an ``\mhmod\ scenario'' which is a 
modification of the \mhmax\ scenario consisting of a reduction of 
$|\Xt/\msusy|$. We define two variants of this scenario, the \mhmodp\
and the
\mhmodm\ scenario, which differ by their sign (and absolute value) of
$\Xt/\msusy$.  
While the positive sign of the product $(\mu\,M_2)$ results in
general in better agreement with the $(g-2)_\mu$ experimental results,
the negative sign of the product ($\mu\,\At$) yields in general 
(assuming minimal flavor violation) better agreement with the 
$\br(b \to s \ga)$ measurements (see \citere{Altmannshofer:2012ks}
for a recent analysis of the impact of other rare $B$ decay observables,
most notably $B_s \to \mu^+\mu^-$). The parameter settings for these two
scenarios are: 

\underline{\mhmodp:}\\[-2em]
\begin{align}
\mt &= 173.2 \gev, \non \\
\msusy &= 1000 \gev, \non \\
\mu &= 200 \gev, \non \\
M_2 &= 200 \gev, \non \\
\Xt^{\OS} &= 1.5\, \msusy  \; \mbox{(FD calculation)}, \non \\
\Xt^{\MSbar} &= 1.6\, \msusy \; \mbox{(RG calculation)}, \non \\ 
\Ab &= \Atau = \At, \non \\
\mgl &= 1500 \gev, \non \\
\msld &= 1000 \gev~.
\label{mhmodp}
\end{align}
\vspace{2em}

\underline{\mhmodm:}\\[-2em]
\begin{align}
\mt &= 173.2 \gev, \non \\
\msusy &= 1000 \gev, \non \\
\mu &= 200 \gev, \non \\
M_2 &= 200 \gev, \non \\
\Xt^{\OS} &= -1.9\, \msusy  \; \mbox{(FD calculation)}, \non \\
\Xt^{\MSbar} &= -2.2\, \msusy \; \mbox{(RG calculation)}, \non \\ 
\Ab &= \Atau = \At, \non \\
\mgl &= 1500 \gev, \non \\
\msld &= 1000 \gev~.
\label{mhmodn}
\end{align}

%%%%%%%%%%%%%%%%%%%%%%%%% F I G U R E %%%%%%%%%%%%%%%%%%%%%%%%%%%%%%%%%%%%%%%%%
\begin{figure}[tb!]
\begin{center}
\includegraphics[width=0.45\textwidth]{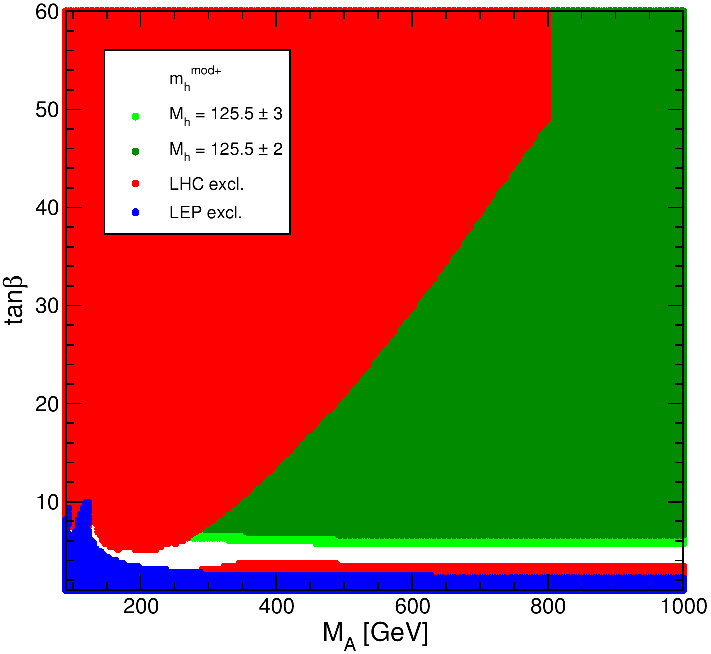}
\includegraphics[width=0.45\textwidth]{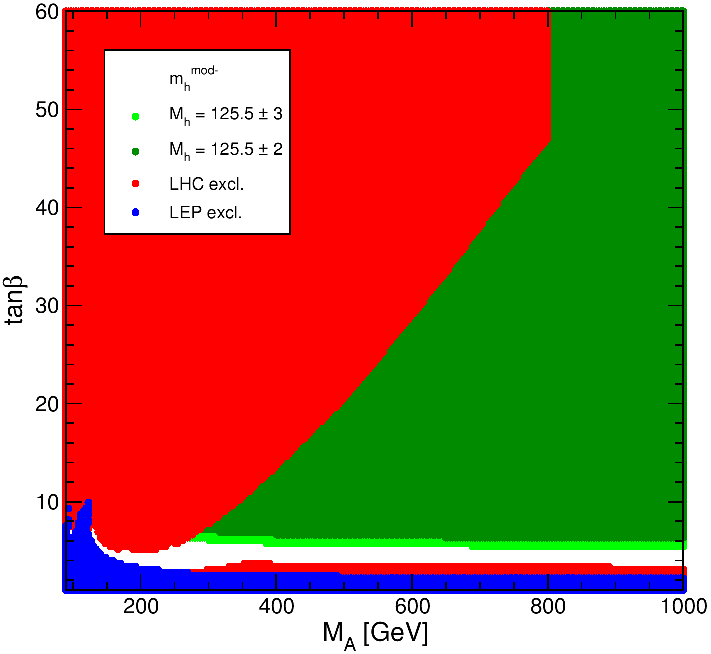}
\caption{
The $\MA$--$\tb$ plane in the \mhmodp\ (left) and \mhmodm\ (right) scenarios.
The colors show exclusion regions from LEP (blue) and the LHC
(red), and the favored region $\Mh=125.5\pm2\, (3)\gev$ (green), see the
text for details. 
}
\label{fig:mhmod}
\end{center}
\vspace{-1em}
\end{figure}
%%%%%%%%%%%%%%%%%%%%%%%%% F I G U R E %%%%%%%%%%%%%%%%%%%%%%%%%%%%%%%%%%%%%%%%%

\begin{figure}[tb!]
\vspace{2em}
\begin{center}
\includegraphics[width=0.45\textwidth]{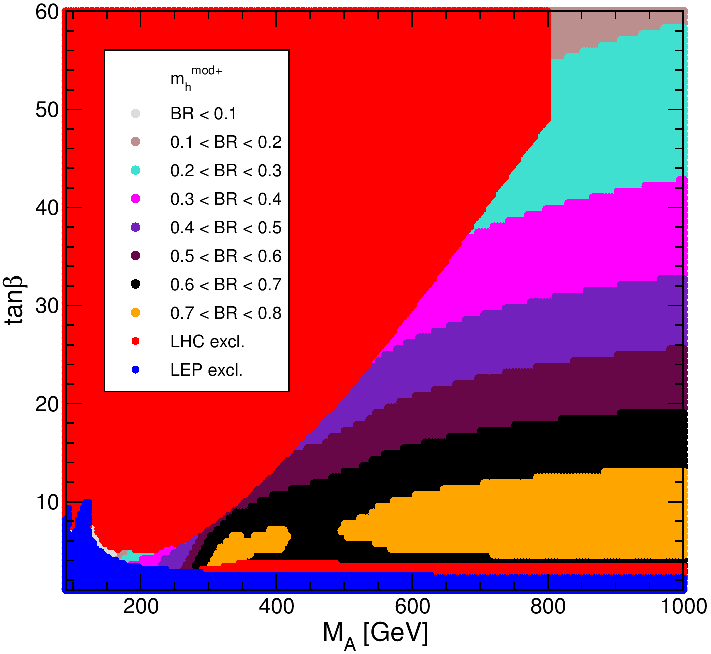}
\includegraphics[width=0.45\textwidth]{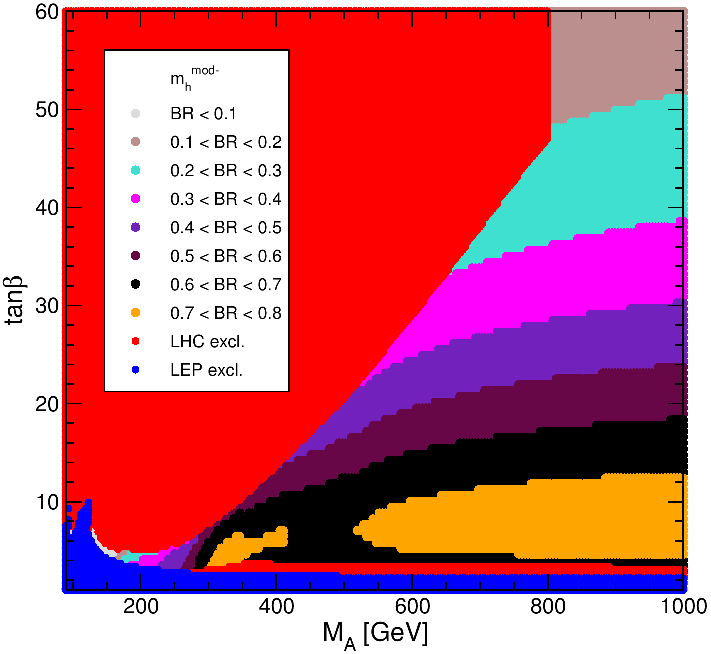}\vspace{1em}
\includegraphics[width=0.45\textwidth]{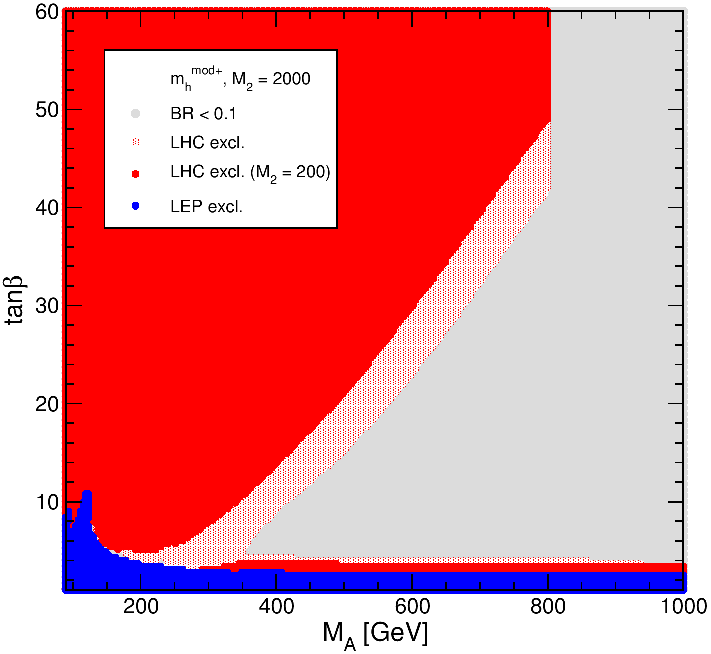}
\includegraphics[width=0.45\textwidth]{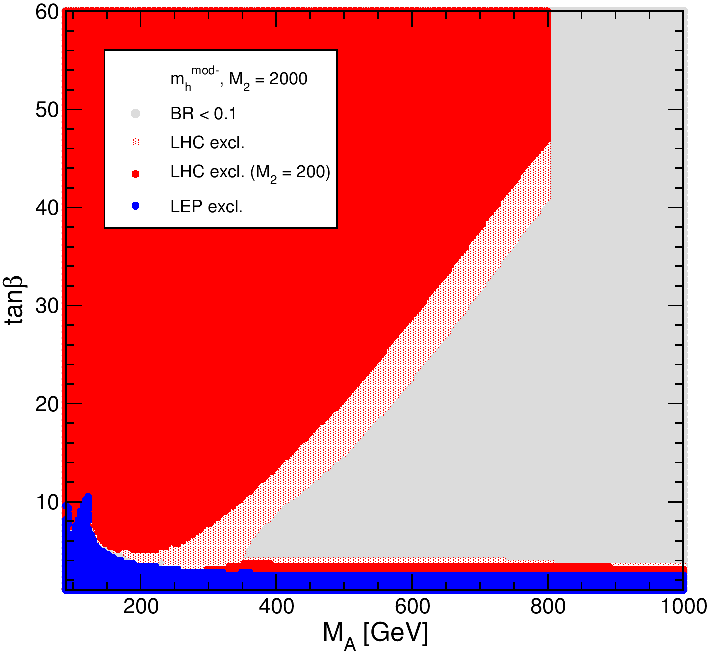}
\caption{Upper row:
The $\MA$--$\tb$ plane in the \mhmodp\ (left) and the \mhmodm\ scenario
(right). The exclusion regions are shown as in \reffi{fig:mhmod},
  while the color coding in the allowed region indicates the average total 
branching ratio of $H$ and $A$ into 
charginos and neutralinos.
In the lower row $M_2 = 2000 \gev$ is used, and the color coding for the
branching ratios of $H$ and $A$ into charginos and neutralinos is as in
the upper row. The regions excluded by the LHC searches are shown in light
red in these plots. For comparison, the excluded regions for the case 
$M_2 = 200\gev$ (as given in the plots in the upper row) is overlaid
(solid red).}
\label{fig:mhmod-BRcn}
\end{center}
\vspace{-1em}
\end{figure}

Figure~\ref{fig:mhmod} shows the bounds on the $\MA$--$\tb$ parameter
space in the \mhmodp\ (left) and \mhmodm\ (right) scenarios, 
using the same choice of colors as in the
\mhmax\ scenario presented in the previous section, but from here
on we show the full LHC exclusion region as solid red only.%
\footnote{The light red color in \reffi{fig:mhmod-BRcn} has a different meaning.
}%
~As 
anticipated, there is
a large region of parameter space at moderate and large values 
of $\tb$ where the mass of the light $\cp$-even Higgs boson 
is in good agreement with the mass value of the
particle recently discovered at the LHC. Accordingly, the green area
indicating the favored region now extends over almost the whole allowed
parameter space of this scenario, with the exception of a small region
at low values of $\tb$. From
\reffi{fig:mhmod} one can see that once the magnitude of $X_t$ has
been changed in order to bring the mass of the light $\cp$-even Higgs
boson into agreement with the observed mass of the signal,
the change of sign of this parameter has a minor impact on the excluded
regions. 

%%%%%%%%%%%%%%%%%%%%%%%%% F I G U R E %%%%%%%%%%%%%%%%%%%%%%%%%%%%%%%%%%%%%%%%%
%%%%%%%%%%%%%%%%%%%%%%%%% F I G U R E %%%%%%%%%%%%%%%%%%%%%%%%%%%%%%%%%%%%%%%%%

As mentioned above, the exclusion limits obtained from the searches
for heavy MSSM Higgs bosons in the $\tau^+\tau^-$ and $b \bar b$ final 
states are significantly affected in parameter regions where additional
decay modes of the heavy MSSM Higgs bosons are open. In particular, 
the branching ratios for the decay of $H$ and $A$ into 
charginos and neutralinos may become large at
small or moderate values of $\tb$, leading to a corresponding reduction
of the branching ratios into $\tau^+\tau^-$ and $b \bar b$. 
In \reffi{fig:mhmod-BRcn} we show again the \mhmodp\ (left) and \mhmodm\
(right) scenarios, where the excluded regions from the Higgs searches at 
LEP and the LHC are as before. In the upper row
of~\reffi{fig:mhmod-BRcn} the color coding for the allowed region of the
parameter space
indicates the average value of the branching ratios for the decay of $H$
and $A$ into charginos and neutralinos (summed over all contributing
final states).% 
\footnote{The branching ratios into charginos and neutralinos 
turn out to be very similar for the heavy $\cp$-even Higgs boson, $H$,
and the $\cp$-odd Higgs boson, $A$, in this region of parameter space.}
One can see from the plots that as a consequence of the
relatively low values of $\mu$ and $M_2$ in this benchmark scenario 
decays of $H$ and $A$ into charginos and neutralinos are kinematically
open essentially in the whole allowed parameter space of the scenario,
with the exception of a small region with rather small $\MA$. The
branching ratios for the decays of $H$ and $A$ into charginos and
neutralinos reach values in excess of 70\% for small and moderate values of
$\tb$.

The impact of the corresponding reduction of the branching ratios
of $H, A$ into $\tau^+\tau^-$ and $b \bar b$ on the excluded region can
be read off from the plots in the lower row of \reffi{fig:mhmod-BRcn}.
In those plots we have set $M_2 = 2000 \gev$, which suppresses the
decays of $H$ and $A$ into charginos and neutralinos. The region
excluded by the LHC searches for MSSM Higgs bosons is shown in light red
for this case. Overlaid for comparison is the excluded region obtained
for $M_2 = 200\gev$, as given by the plots in the upper row (solid red). 
One can see that the impact of the decays into charginos and neutralinos 
on the excluded region in the $\MA$--$\tb$ plane is sizable, amounting
typically to a shift in the excluded value for $\tb$ by more than 
$\De\tb = 5$ for a given value of $\MA$.

As mentioned above, another decay mode that is kinematically
possible over a large part of the parameter space of the \mhmod\ 
scenarios is the decay rate of
$H \to hh$. For $M_2 = 200\gev$ (plots in the
upper row of \reffi{fig:mhmod-BRcn}) and $\MA = 300 \gev$ we find in the 
\mhmodp\ (\mhmodm) scenario
$\br(H \to hh) = 12\%$ $(11\%)$ for $\tb = 7$ and
$\br(H \to hh) = 17\%$ $(16\%)$ for $\tb = 6$. Increasing $M_2$ to 
$M_2 = 2000\gev$ (plots in the
lower row of \reffi{fig:mhmod-BRcn}) suppresses the 
decays into charginos and neutralinos, and correspondingly enhances the decay $H \to hh$. For
$\MA = 300 \gev$ in the
\mhmodp\ (\mhmodm) scenario we obtain
$\br(H \to hh) = 19\%$ $(18\%)$ for $\tb = 7$ and
$\br(H \to hh) = 29\%$ $(27\%)$ for $\tb = 6$. 
As already mentioned, we encourage ATLAS and CMS to enhance the 
sensitivity of their searches
for MSSM Higgs bosons by performing also dedicated searches for Higgs
decays into SUSY particles 
and into a pair of lighter Higgs bosons.

%%%%%%%%%%%%%%%%%%%%%%%%% F I G U R E %%%%%%%%%%%%%%%%%%%%%%%%%%%%%%%%%%%%%%%%%
\begin{figure}[t!]
\begin{center}
\includegraphics[width=0.45\textwidth]{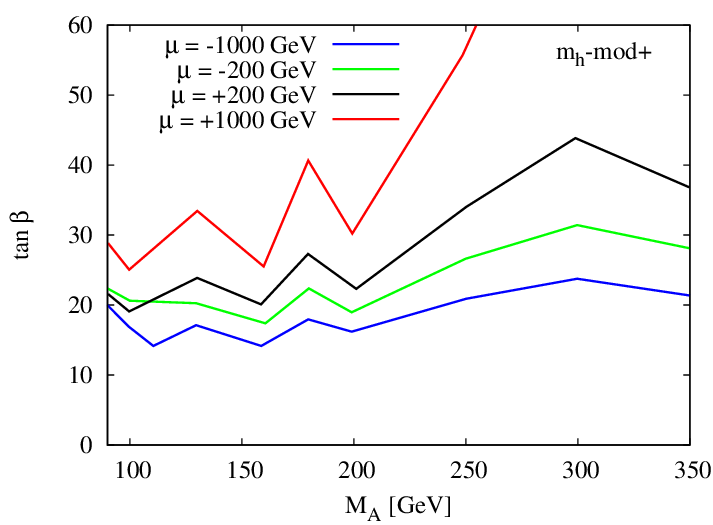}
\includegraphics[width=0.45\textwidth]{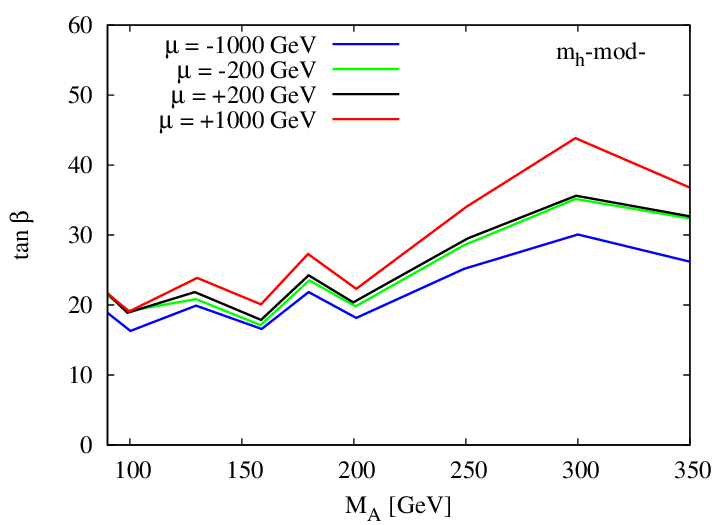}
\caption{
Exclusion limits from the most recent CMS
analysis of the channel $b \bar b \phi, \phi \to b \bar b$ 
(with $\phi= h, H,A$)~\cite{CMSbbHbb} are presented in the $\MA$--$\tb$
plane for the scenarios \mhmodp\ (left) and \mhmodm\ (right) with
variation of the $\mu$ parameter as indicated by the legend. 
}
\label{fig:mhmod-bbHAbb}
\end{center}
%\vspace{2em}
\end{figure}
%%%%%%%%%%%%%%%%%%%%%%%%% F I G U R E %%%%%%%%%%%%%%%%%%%%%%%%%%%%%%%%%%%%%%%%%

For the benchmarks proposed in this paper a certain value for the
parameter $\mu$ is specified. However, we suggest to investigate the impact of 
an enhancement or suppression of the bottom Yukawa coupling by varying
the parameter $\mu$ according to \refeq{eq:musuggest}. For the
Higgs decays into $\tau^+\tau^-$, see \refeq{phi-tautau}, a partial
cancellation of the associated $\db$ corrections occurs between the 
contributions to the production and the decay, leading to a relatively
mild dependence on the bottom Yukawa coupling and therefore on
$\db$~\cite{benchmark3}. On the other hand, for the associated
production and decay into bottom quarks, see \refeq{bb-phi}, the 
$\db$ corrections enter in a similar way for the production and decay
part, so that their overall effect is significantly larger, leading to a
more pronounced dependence on the sign and size of the $\mu$
parameter~\cite{benchmark3}. Negative values of $\mu$
lead to a stronger bottom-quark Yukawa coupling and therefore a larger
production rate and a larger parameter range exclusion. 
The bounds on the parameter space from this channel tend
to be weaker than those from $\tau\tau$ searches, and they
are therefore not explicitly visible in \reffi{fig:mhmod}.
In order to display the effect of the corrections to the bottom Yukawa
coupling we focus now explicitly on the 
channel $b \bar b \phi,  \phi \to b \bar b$, where $\phi = h,H, A$.
Using the latest result from CMS for this channel~\cite{CMSbbHbb},
\reffi{fig:mhmod-bbHAbb} shows 
the reach in the $\MA$--$\tb$ plane of the \mhmodp\ (left) and
\mhmodm\ (right) scenarios for 
$\mu = \pm 200 \gev, \pm 1000\gev$ (see also \cite{Carena:2012rw}).%
\footnote{We have verified our implementation of this
  limit against the results from CMS~\cite{CMSbbHbb}, which are given
  for the (original) \mhmax\ scenario with $\mu=\pm 200\gev$. The
  ``zig-zag''-type variation 
  of the bounds originates from the original bounds in \citere{CMSbbHbb}.}
In the \mhmodp\ scenario one can observe a very large variation
with the sign and absolute value of $\mu$. For example, for 
$\MA = 250\gev$ one finds for $\mu = - 1000\gev$ an exclusion in $\tb$
down to about $\tb = 20$, while for the reversed sign of $\mu$ the
excluded region starts only above $\tb = 50$. The dependence on $\mu$ is less
pronounced in the \mhmodm\ scenario, i.e.\ for negative values of $X_t$,
which is a consequence of a 
partial compensation between the main contributions to $\db$,
see \refeq{def:dmb}.

%%%%%%%%%%%%%%%%%%%%%%%%%%%%%%%%%%%%%%%%%%%%%%%%%%%%%%%%%%%%%%%%%%%%%%%%%%%%%%%

%\clearpage
%\newpage
\subsection{The light stop scenario}
\label{sec:lightstop}

The measured value of the lightest $\cp$-even Higgs mass of about $\mass\gev$
may only be achieved in the MSSM by relatively large radiative contributions
from the top--stop sector.  It is well known that this can only be obtained if
the mixing parameter $X_t$ in the stop sector is larger than the average
stop mass. The dependence of $\Mh$ on the stop mass scale is logarithmic and
allows for values of $M_{\rm SUSY}$ below the TeV scale. Values 
of $\msusy$
significantly below the TeV scale are still possible if $X_t$ is
close to the value that maximizes the lightest $\cp$-even Higgs mass
(or, to a lesser extent, close to the maximum for negative values of
$X_t$). Such a large value of $|X_t|$ and a relatively low value 
of $M_{\rm SUSY}$ necessarily
lead to the presence of a light stop. Such a light stop may be searched
for in direct production at the LHC, but has also a relevant impact on
the lightest $\cp$-even Higgs production rates. In particular, 
a light stop may lead to a relevant modification of the gluon fusion
rate~\cite{ggh-djouadi,benchmark2}. 

The contribution of light stops to the gluon fusion amplitude may be
parametrized in terms of the physical stop masses and the mixing
parameter. 
Making use of low energy theorems~\cite{Lowenergy} it is easy to
see that the stops give rise 
to an additional contribution to the gluon fusion amplitude which is
approximately given by~\cite{Stopsglueglue} 
\begin{equation}
\delta {\cal{A}}_{hgg}/{\cal{A}}_{hgg}^{\rm SM} \simeq  
       \frac{m_t^2}{4 m_{\tilde{t}_1}^2 m_{\tilde{t}_2}^2}
\left( m_{\tilde{t}_1}^2 + m_{\tilde{t}_2}^2 - X_t^2 \right)~,   
\label{glufusion}
\end{equation}
where ${\cal{A}}_{hgg}^{\rm SM}$ denotes the gluon fusion amplitude in
the SM.
Values of $X_t$ in the range $2M_{\rm SUSY} \lesssim X_t \lesssim
2.5M_{\rm SUSY} $ then lead to negative contributions to this amplitude 
and to reduced values of the gluon fusion rate.  We propose a
\lstop\ scenario with the following parameters,

\underline{\lstop:}\\[-2em]
\begin{align}
\mt &= 173.2 \gev, \non \\
\msusy &= 500 \gev, \non \\
\mu &= 350 \gev, \non \\
M_2 &= 350 \gev, \non \\
\Xt^{\OS} &= 2.0\, \msusy  \; \mbox{(FD calculation)}, \non \\
\Xt^{\MSbar} &= 2.2\, \msusy \; \mbox{(RG calculation)}, \non \\ 
\Ab &= \At = \Atau, \non \\
\mgl &= 1500 \gev, \non \\
\msld &= 1000 \gev~.
\label{lightstop}
\end{align}

These parameters
lead to a lighter stop and a heavier stop mass of about 
$325 \gev$ and $670 \gev$, 
respectively, and a negative correction of the gluon fusion
amplitude of about 8\%. The \lstop\ scenario can be regarded as an
update of the \gluophobic\ scenario defined in \citere{benchmark2}.

The values of $\mu$ and $M_2$ in the \lstop\ scenario have been
chosen to be in agreement with the current
exclusion bounds on direct light stop production at the
LHC~\cite{Stopsearches}.%
\footnote{
The values of $\mu$, $M_1$ and $M_2$
could be adjusted to slightly larger values
if the currently proposed values were excluded by future experiments. 
For instance, the choice $M_1 = 350 \gev$, $M_2 = \mu = 400 \gev$ leads
to a SUSY spectrum that is very difficult to test at the LHC.
In general, for a given value of $\tb$ and $\MA$, slightly larger values of
$\mu$ and $M_{1,2}$ 
would lead to a small decrease of the value of $\Mh$ and therefore to a small
shift of the green areas to larger values of 
$\tb$.}
 The two-body decay modes that are kinematically open are $\Stope
\to b \chap{1}$ and  
$\Stope \to c \neu{1}$ with $\mcha{1} \approx 295 \gev$ and 
$\mneu{1} \approx 163 \gev$. The first decay results
in very soft decay products. While the latter decay is expected to be
suppressed in minimal flavor violating schemes, it could in general be
sizable.
Analyses have been performed at the Tevatron~\cite{StopTev}; however,
currently there are no dedicated LHC searches in this channel. If 
this channel turned out to be relevant, due to its difficult final
state it would pose a challenge to the experimental analyses.

There is also a correction to the diphoton amplitude, but since in 
the diphoton case the dominant SM contribution comes from $W$ loops, 
which are of opposite sign and about a factor $4$ larger than the top
contributions, the stop contributions lead to only a small modification,
smaller than about 3\%, of this amplitude.  

%%%%%%%%%%%%%%%%%%%%%%%%% F I G U R E %%%%%%%%%%%%%%%%%%%%%%%%%%%%%%%%%%%%%%%%%
\begin{figure}[t!]
\begin{center}
\includegraphics[width=0.45\textwidth]{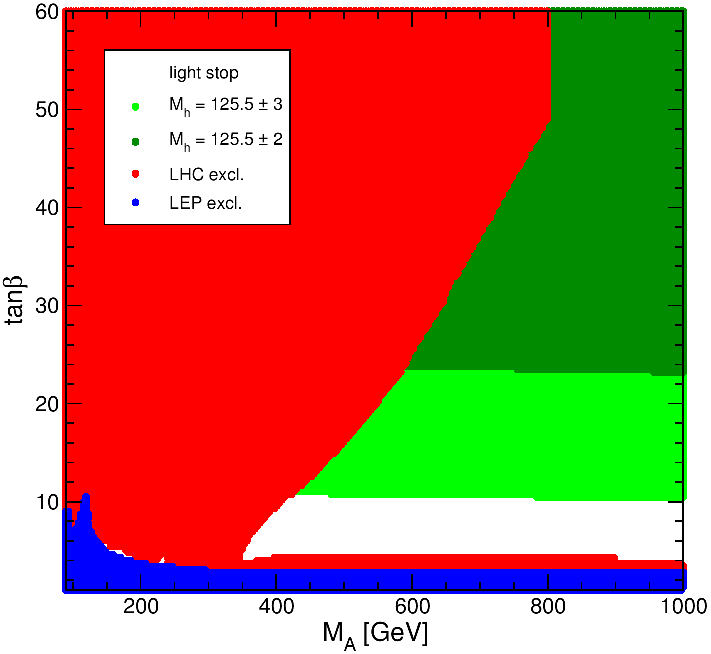}
\includegraphics[width=0.45\textwidth]{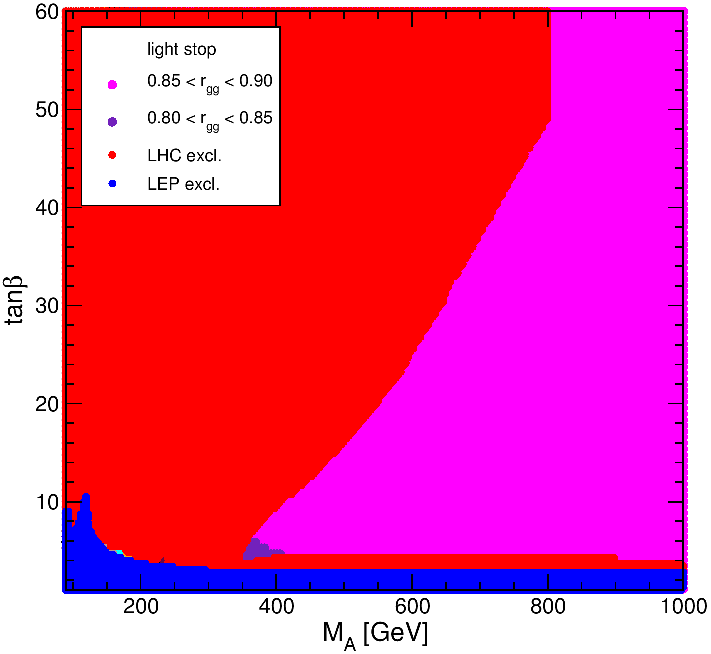}
\caption{
The $\MA$--$\tb$ plane in the \lstop\ scenario; left: with the same
color coding as in \reffi{fig:mhmod}; 
right: the resulting suppression of 
the gluon fusion rate, as indicated by the legend. 
}
\label{fig:lightstop}
\end{center}
%\vspace{2em}
\end{figure}
%%%%%%%%%%%%%%%%%%%%%%%%% F I G U R E %%%%%%%%%%%%%%%%%%%%%%%%%%%%%%%%%%%%%%%%%

Figure~\ref{fig:lightstop} shows the $\MA$--$\tb$ plane in the
\lstop\ scenario, as well as a comparison of the gluon fusion rates for
$h$ production to those obtained in the SM. For this comparison, we
define the quantity
\begin{equation}
r_{gg}=\frac{\Ga(h\to gg)_{\mathrm{MSSM}}}
            {\Ga(h\to gg)_{\mathrm{SM}}}~,
\label{eq:Rgg}
\end{equation}
which gives a rough approximation of the relative suppression of
$\si(gg \to h)_{\rm MSSM}$.
The bounds on the parameter space (as before obtained with 
{\tt HiggsBounds}) are similar to the ones
obtained in the \mhmod\ scenarios. However, the gluon fusion rate is
between 10\% and 15\% lower than in the SM, as expected from
\refeq{glufusion}. This shift is similar in magnitude to the 
current theoretical uncertainties on the gluon fusion cross section from
e.g.~the strong coupling constant and parton distribution functions.

%%%%%%%%%%%%%%%%%%%%%%%%%%%%%%%%%%%%%%%%%%%%%%%%%%%%%%%%%%%%%%%%%%%%%%%%%%%%%%%

%\newpage
\subsection{The light stau scenario}

While light stops may lead to a large modification of the gluon fusion
rate, with a relative minor effect on the diphoton rate, it has been
shown that light staus, in the presence of large mixing, may lead to
important modifications of the diphoton decay width  of the lightest
$\cp$-even Higgs boson, 
$\Ga(h \to \ga\ga)$~\cite{LightStau1,LightStau2}. Large
mixing in the stau sector may happen naturally for large values of
$\tb$, for which the mixing parameter $X_\tau = A_\tau - \mu \tb$ 
becomes large. Similarly to the modifications of the gluon
fusion rate in the \lstop\ scenario, one can use the low energy
Higgs theorems~\cite{Lowenergy} to obtain the modifications of the
decay rate of the Higgs boson to photon pairs. 
The correction to the amplitude of Higgs decays to diphotons  
is approximately given by~\cite{LightStau1,Stopsglueglue} 
\begin{equation}
\de{\cal{A}}_{h\gamma\gamma} / {\cal{A}}_{h\gamma\gamma}^{\rm SM} \simeq  
             -\frac{2 \; m_\tau^2}{39 \; m_{\tilde{\tau}_1}^2 m_{\tilde{\tau}_2}^2}
\left( m_{\tilde{\tau}_1}^2 + m_{\tilde{\tau}_2}^2 - X_\tau^2 \right) ,
\label{diphoton}
\end{equation}
where ${\cal{A}}_{h\ga\ga}^{\rm SM}$ denotes the diphoton
amplitude in the SM.

Due to the large $\tb$ enhancement $X_\tau$ is naturally much
larger than the stau masses and hence the corrections are positive and
become significant for large values of $\tb$. 
As stressed above, the current central value of the
measured diphoton rate of the state discovered at the LHC is
somewhat larger than the expectations for a SM Higgs, 
which adds motivation for investigating the phenomenology of a scenario
with an enhanced diphoton rate. We therefore propose a \lstau\ scenario. 
In the definition of the parameters we distinguish the cases whether or
not $\tau$ mass threshold corrections,
$\Delta_\tau$, are incorporated in the computation of the stau spectrum
(this is the case in {\tt CPsuperH}, but not in the present version of 
{\tt FeynHiggs}). We mark the case where those corrections are included
as ``($\De_\tau$ calculation)''. We define the parameters of the 
\lstau\ scenario as follows: 

%\clearpage
%\newpage

\underline{\lstau:}\\[-2em]
\begin{align}
\mt &= 173.2 \gev, \non \\
\msusy &= 1000 \gev, \non \\
\mu &= 500 \gev, \non \\
\mu &= 450 \gev \; (\De_\tau \mbox{~calculation}), \non \\
M_2 &= 200 \gev, \non \\
M_2 &= 400 \gev \; (\De_\tau \mbox{~calculation}), \non \\
\Xt^{\OS} &= 1.6\, \msusy  \; \mbox{(FD calculation)}, \non \\
\Xt^{\MSbar} &= 1.7\, \msusy \; \mbox{(RG calculation)}, \non \\
\Ab &= \At ~, \non \\
\Atau &= 0~, \non \\
\mgl &= 1500 \gev, \non \\
\msld &= 245 \gev, \non \\
\msld &= 250 \gev \; (\De_\tau \mbox{~calculation}) .
\label{lightstau}
\end{align}

\begin{figure}[htb!]
\begin{center}
\includegraphics[width=0.45\textwidth]{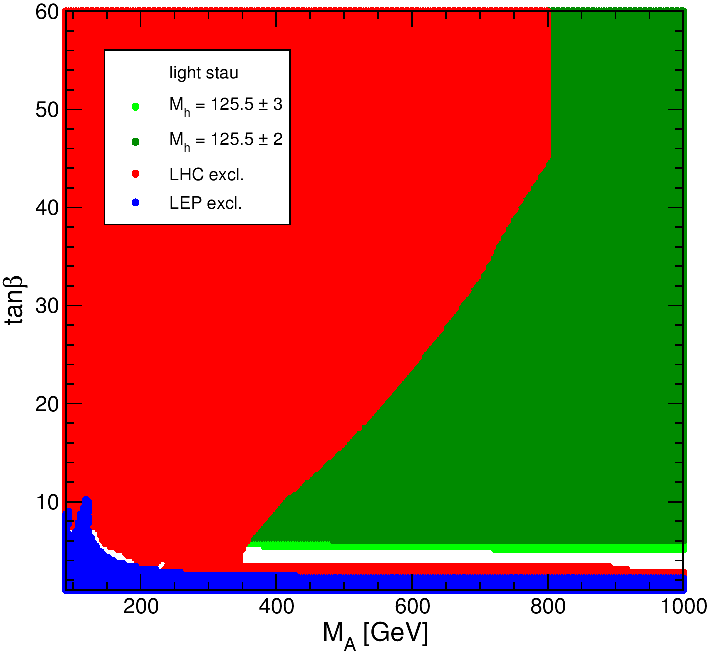}
\includegraphics[width=0.45\textwidth]{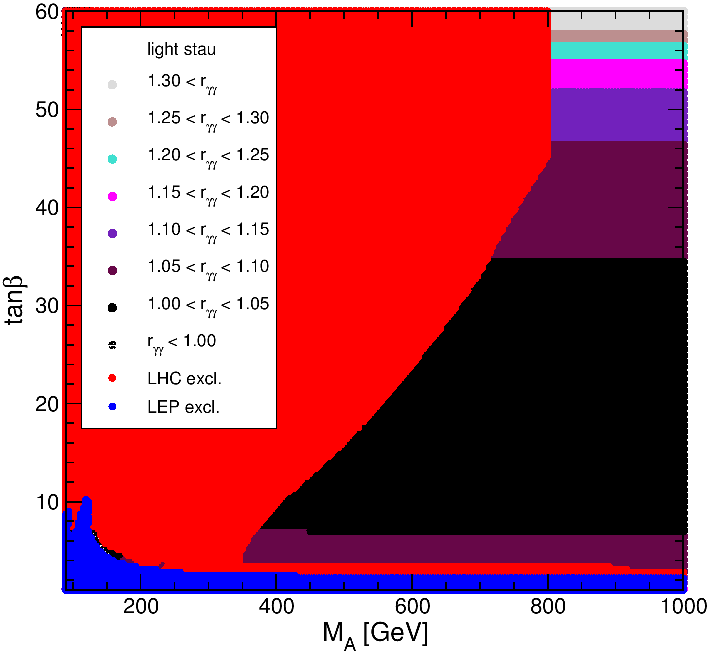}
\caption{Left: The $\MA$--$\tb$ plane in the \lstau\ scenario, with the
  same color coding as in \reffi{fig:mhmod}.
Right: The effect of light staus on the decay rate $h \to \ga\ga$,
where the quantity $r_{\gamma\gamma}$ is defined in analogy to 
$r_{gg}$ in~\refeq{eq:Rgg}.
}
\label{fig:lightstau}
\end{center}
%\vspace{2em}
\end{figure}

Figure~\ref{fig:lightstau} shows the $\MA$--$\tb$
plane in the \lstau\ scenario (left), as well as comparison of the 
$h \to \ga\ga$ width to the SM case (right). 
Concerning the exclusion bounds
from the Higgs searches at LEP and the LHC, the main 
difference with respect to the \mhmod\ scenarios is present
at low values of $\tb$, where the LHC exclusion in the  \lstau\ scenario
is somewhat stronger. This results from a suppression of the decays into 
charginos and neutralinos caused by the relatively large (default) 
value of $\mu$ in
the \lstau\ scenario. The right panel shows the enhancement of the
diphoton decay rate of the lightest $\cp$-even Higgs boson with respect 
to the SM (with $r_{\gamma\gamma}$ defined analogously
to $r_{gg}$ in~\refeq{eq:Rgg}). As 
expected, a significant enhancement is present at large values of
$\tb > 50$, for which the lightest stau approaches a mass of about
100~GeV, close to the LEP limit for the stau mass~\cite{pdg}.
For non-zero values of $A_\tau$ in this scenario, the coupling of the
down-type fermions to the lightest Higgs boson may be modified
\cite{LightStau1}. The decay rate of $H/A$ into staus can also become
sizable, see the discussion in Sect.~\ref{sect:tauphobic}. 

%%%%%%%%%%%%%%%%%%%%%%%%% F I G U R E %%%%%%%%%%%%%%%%%%%%%%%%%%%%%%%%%%%%%%%%%
%%%%%%%%%%%%%%%%%%%%%%%%% F I G U R E %%%%%%%%%%%%%%%%%%%%%%%%%%%%%%%%%%%%%%%%%

%%%%%%%%%%%%%%%%%%%%%%%%%%%%%%%%%%%%%%%%%%%%%%%%%%%%%%%%%%%%%%%%%%%%%%%%%%%%%%%

\newpage 
\subsection{The \boldmath{$\tau$}-phobic Higgs scenario}
\label{sect:tauphobic}
Besides the loop effects on the Higgs vertices described in the previous
sections, also propaga- tor-type corrections involving the mixing
between the two $\cp$-even Higgs bosons of the MSSM can have an
important impact. In particular, this type of corrections can lead to
relevant modifications of the Higgs couplings to down-type fermions,
which can approximately be taken into account via an effective mixing
angle~$\aeff$ (see
\citere{hff}). This modification occurs for large values of the  
$A_{t,b,\tau}$ parameters and large values of $\mu$ and $\tb$.%
\footnote{
Large values of $A_{t,b,\tau}$ and $\mu$ are in principle constrained by the requirement that no
charge and color breaking minima should appear in the potential~\cite{ccb}, or at least that there is a sufficiently long-lived meta-stable vacuum. However, a detailed analysis of this issue is beyond the scope of this paper, and we leave it for a future analysis.
%Large values of $A_{t,b,\tau}$ and $\mu$ can in principle
%be problematic as they can lead to
%charge and color breaking minima in the potential~\cite{ccb}. However,
%the values chosen in the \tauphobic\ scenario
%should yield at least a sufficiently meta-stable
%vacuum.
}
  
The scenario that we propose can be regarded as an update of the
\saeff\ scenario proposed in \citere{benchmark2}. The parameters are:

\underline{\tauphobic:}\\[-2em]
\begin{align}
\mt &= 173.2 \gev, \non \\
\msusy &= 1500 \gev, \non \\
\mu &= 2000 \gev, \non \\
M_2 &= 200 \gev, \non \\
\Xt^{\OS} &= 2.45\,\msusy  \; \mbox{(FD calculation)}, \non \\
\Xt^{\MSbar} &= 2.9\,\msusy \; \mbox{(RG calculation)}, \non \\ 
\Ab &= \Atau = \At ~, \non \\
\mgl &= 1500 \gev, \non \\
\msld &= 500 \gev ~.
\label{tauphobic}
\end{align}

The relatively low value of $\msld = 500 \gev$ and the large value
of $\mu$ give rise to rather light staus also in the \tauphobic\
scenario, in particular in the region of
large $\tb$. The corrections from the stau sector have an important
%\mpar{check statements in this paragraph!\\ Expand?}
influence on the Higgs couplings to down-type fermions in this scenario. 
Furthermore, in this scenario decays of the heavy $\cp$-even Higgs boson
into light staus, $H \to \tilde\tau_1^+\tilde\tau_1^-$, occur with a
large branching fraction in the region of large $\tb$ and sufficiently
high $\MA$. For example, for $\MA = 800 \gev$ and $\tb = 45$, we obtain
$\br(H\to \tilde\tau_1^+\tilde\tau_1^-) = 67\%$. 

%%%%%%%%%%%%%%%%%%%%%%%%% F I G U R E %%%%%%%%%%%%%%%%%%%%%%%%%%%%%%%%%%%%%%%%%
\begin{figure}[t!]
\vspace{2em}
\begin{center}
\includegraphics[width=0.45\textwidth]{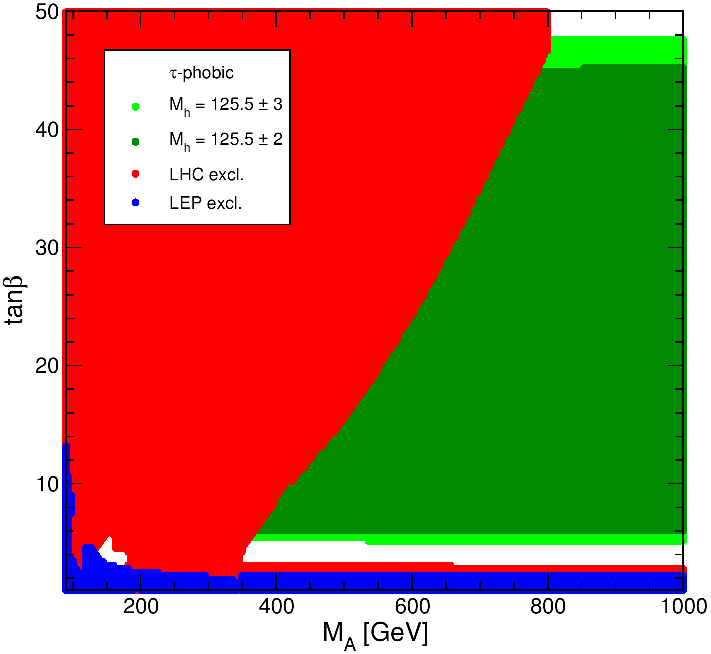}\\
\caption{
The $\MA$--$\tb$ plane in the \tauphobic\ scenario.
The color coding is the same as in \reffi{fig:mhmod}.}
\label{fig:tauphobic}
\end{center}
%\vspace{2em}
\end{figure}
%%%%%%%%%%%%%%%%%%%%%%%%% F I G U R E %%%%%%%%%%%%%%%%%%%%%%%%%%%%%%%%%%%%%%%%%

Figure~\ref{fig:tauphobic} shows the bounds on the $\MA$--$\tb$
parameter space in the \tauphobic\ scenario. 
As in the \lstau\ scenario, the most
important modification with respect to the \mhmod\ scenarios is a larger
exclusion at low values of $\tb$ induced by a decrease of the
decay rate into charginos and neutralinos. 

%%%%%%%%%%%%%%%%%%%%%%%%% F I G U R E %%%%%%%%%%%%%%%%%%%%%%%%%%%%%%%%%%%%%%%%%
\begin{figure}[tb!]
\begin{center}
\includegraphics[width=0.45\textwidth]{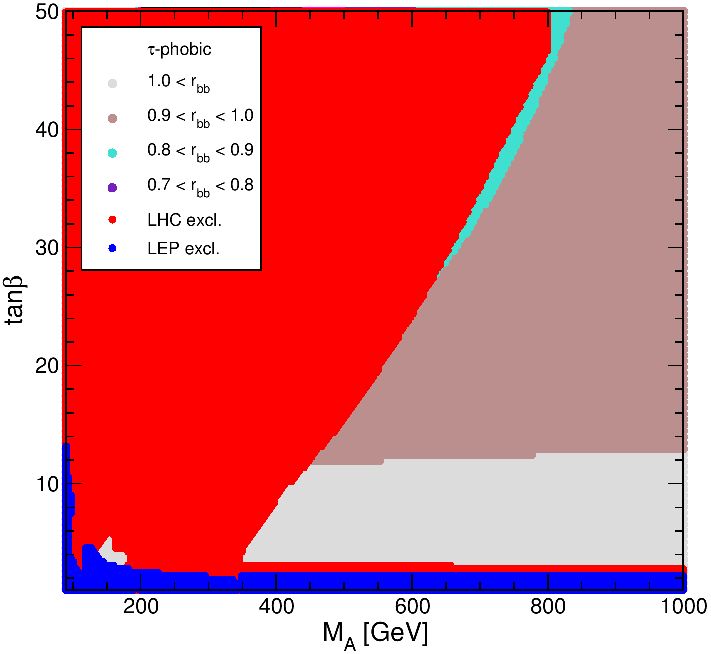}
\includegraphics[width=0.45\textwidth]{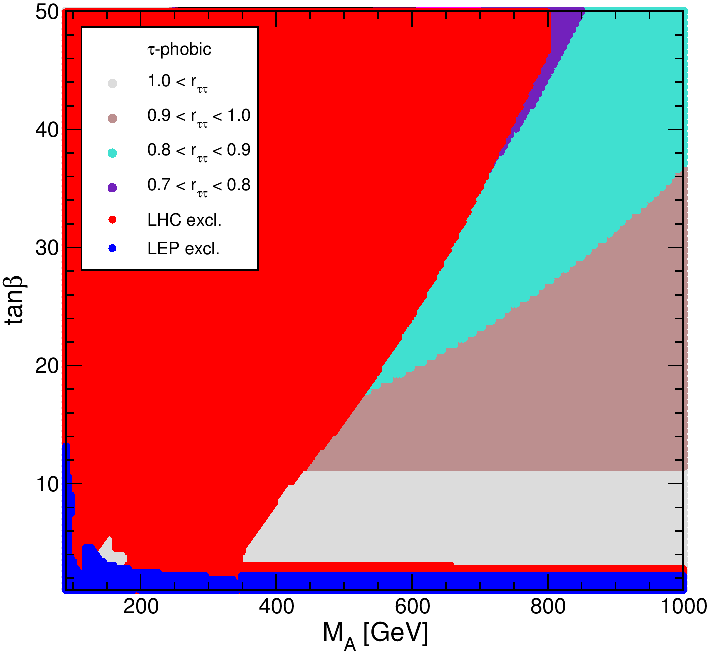}
\caption{
Modification of the decay rate for the
lightest $\cp$-even Higgs boson into bottom quarks ($r_{bb}$, left) and
$\tau$-leptons ($r_{\tau\tau}$, right) in the \tauphobic\ scenario, where 
$r_{bb}$ and $r_{\tau\tau}$ are defined in analogy to $r_{gg}$ in 
\refeq{eq:Rgg}.
}
\label{fig:tauphobic_R}
\end{center}
\vspace{-1em}
\end{figure}
%%%%%%%%%%%%%%%%%%%%%%%%% F I G U R E %%%%%%%%%%%%%%%%%%%%%%%%%%%%%%%%%%%%%%%%%

Figure~\ref{fig:tauphobic_R}
shows the modification of the decay rate for the
lightest $\cp$-even Higgs boson into bottom quarks ($r_{bb}$) and
$\tau$-leptons ($r_{\tau\tau}$), both defined analogously to $r_{gg}$,
see \refeq{eq:Rgg}.
The variations are most important at large values of $\tb$,
and they increase
for smaller values of $\MA$, where the LHC exclusion limit from MSSM
Higgs searches becomes very significant. Still, as can be seen from the
figure, modifications of the partial Higgs decay width into $\tau^+\tau^-$
larger than 20\%,
and of the decay width into bottom quarks larger than
10\% may occur within this scenario.

%%%%%%%%%%%%%%%%%%%%%%%%%%%%%%%%%%%%%%%%%%%%%%%%%%%%%%%%%%%%%%%%%%%%%%%%

\newpage
\subsection{The low-\boldmath{$\MH$} scenario}

As it was pointed out in \citeres{Mh125,NMSSMLoopProcs,MH125other},
besides the interpretation of the Higgs-like state at $\sim \mass \gev$
in terms of the light $\cp$-even Higgs boson of the MSSM it is also
possible, at least in principle, to identify the observed signal 
with the {\em heavy} $\cp$-even Higgs boson of the MSSM. In this case
the Higgs sector would be very different from the SM case, since all 
five MSSM Higgs bosons would be light. The heavy $\cp$-even Higgs boson 
would have a mass around $\mass \gev$ and
behave roughly SM-like, while the light $\cp$-even Higgs boson of the
MSSM would have heavily suppressed couplings to gauge bosons. Due to the
rather spectacular phenomenology of such a scenario, the available
parameter space is already affected by existing search limits, and the 
prospects for discovering a non-SM like Higgs in the near future would
be very good. 

The most relevant limits probing such a scenario at present arise from the 
searches for MSSM Higgs bosons in the $gg,b\bar b \to h, H, A \to \tau\tau$
channel,
but also the search for a light charged Higgs in top quark decays
has an interesting sensitivity. The results for the 
$gg,b\bar b \to h, H, A \to \tau\tau$ channel have recently been updated by
CMS~\cite{HCP2012tautau}. However, it is difficult to assess the impact of
those new results on the viability of such a scenario, since they have
been presented only for the \mhmax\ scenario (i.e., no cross section
limits have been provided which could readily be applied to other
scenarios; an attempt to incorporate a rough estimate of the new CMS
result has been made in 
{\tt HiggsBounds\,4.0.0}~\cite{higgsbounds,HBmanual}, which we have used
for producing the plots in this paper).
Besides Higgs search limits also limits from flavor physics can place
relevant constraints on this kind of scenario. 
It was found in \citeres{NMSSMLoopProcs,ADNM} that flavor constraints could
lead to tension with the allowed parameter space (which might be aleviated by
taking into account some Non-Minimal Flavor Violation~\cite{jillana}). 
We do not take these indirect constraints into account in this analysis.
In view of the rich and interesting phenomenology, we include a scenario of
this kind among the benchmarks that we 
propose. In particular, this scenario could provide a useful benchmark
for the ongoing charged Higgs boson searches in the MSSM.

In this scenario we deviate from the definition of an $\MA$--$\tb$ plane,
since it is clear that a relatively small value of $\MA$ (and
correspondingly $\MHp$) is required.
$\MA$ is therefore fixed to $\MA = 110\gev$ (other choices for
$\MA$ in this low-mass region would also be possible),
and instead $\mu$ is varied. Otherwise we choose the same parameters as
for the \tauphobic\ scenario, with the exception that we set 
$\msld = 1000 \gev$, while the value in the \tauphobic\ scenario is
$\msld = 500\gev$ (see the discussion above). Accordingly, the
parameters proposed for this scenario are:%
\footnote{
The remark made in the previous section about the constraints from 
charge and color breaking minima in the scalar potential applies also here.
}

\underline{\lowMH:}\\[-2em]
\begin{align}
\mt &= 173.2 \gev, \non \\
\MA &= 110 \gev, \non \\
\msusy &= 1500 \gev, \non \\
M_2 &= 200 \gev, \non \\
\Xt^{\OS} &= 2.45\, \msusy  \; \mbox{(FD calculation)}, \non \\
\Xt^{\MSbar} &= 2.9\, \msusy \; \mbox{(RG calculation)}, \non \\ 
\Ab &= \Atau = \At, \non \\
\mgl &= 1500 \gev, \non \\
\msld &= 1000 \gev~.
\label{lowMH}
\end{align}
Instead of $\MA$ one can also use $\MHp$ as input parameter, as it
is done, e.g., in {\tt CPsuperH}. In this case one should choose
as input value $\MHp = 132 \gev$, leading to very similar phenomenology.

%%%%%%%%%%%%%%%%%%%%%%%%% F I G U R E %%%%%%%%%%%%%%%%%%%%%%%%%%%%%%%%%%%%%%%%%
\begin{figure}[htb!]
\vspace{-1em}
\begin{center}
\includegraphics[width=0.55\textwidth]{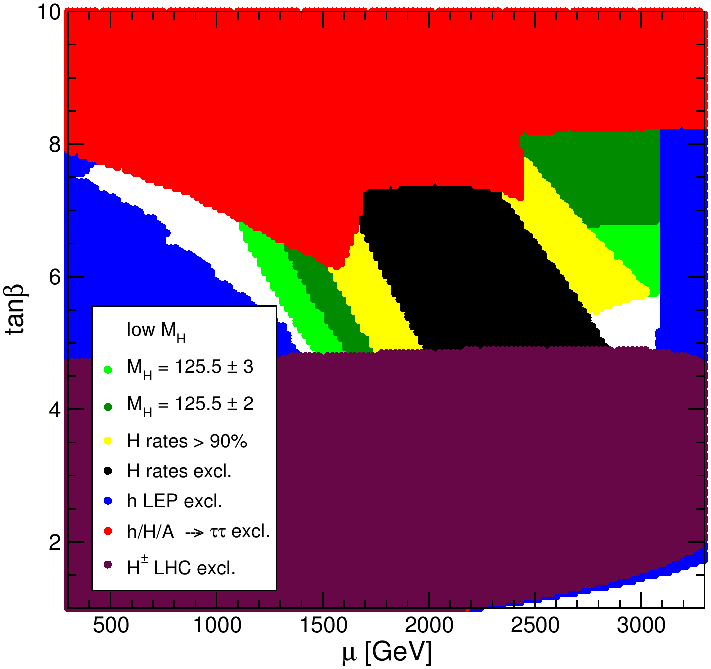}
\caption{
Experimentally favored and excluded regions in the $\mu$--$\tb$ plane in
the \lowMH\ scenario. Details of the color coding (as indicated in the 
legend) are
described in the text.} 
\label{fig:lowMH}
\end{center}
%\vspace{-2em}
\end{figure}
%%%%%%%%%%%%%%%%%%%%%%%%% F I G U R E %%%%%%%%%%%%%%%%%%%%%%%%%%%%%%%%%%%%%%%%%

In \reffi{fig:lowMH} we show the $\mu$--$\tb$ plane in the
\lowMH\ scenario. The green shades indicate the region where
$\MH=125.5\pm 2\, (3) \gev$. The yellow and black areas also have
$\MH = \mass \pm 3 \gev$, where the yellow area additionally satisfies the
requirement that the rates for the $gg \to H$, $H\to \ga\ga$ 
and $H \to ZZ^{*}$ channels, as approximated by ($X = \ga, Z$)
\begin{equation}
R_{XX}=\frac{\Ga(H\to gg)_{\MSSM}\times \br(H\to XX)_{\MSSM}}
           {\Ga(H\to gg)_{\SM}\times \br(H\to XX)_{\SM}}~,
\end{equation}
are at least at 90\% of their SM
value for the same Higgs mass. The black region in \reffi{fig:lowMH}
indicates where the rates for $H$ decay to gauge bosons become too high,
such that these points are excluded by {\tt HiggsBounds}. 
As before, the blue area is excluded by LEP Higgs searches, whereas 
the solid red is excluded from LHC searches for the
neutral MSSM Higgs bosons, $h$, $H$ and $A$ in the $\tau^+\tau^-$ decay
channel.
The purple region is excluded
by charged Higgs boson searches at the LHC. The white area at very large
values of $\mu$ and low $\tb$ is unphysical, i.e.\ this parameter
region is theoretically inaccessible. 

One can see from Fig.~\ref{fig:lowMH} that, as expected, such a scenario is
confined to a relatively small range of $\tb$ values (and, as discussed
above, the same holds for $\MA$). It is interesting to note that the
searches for all five MSSM Higgs bosons contribute in a significant way
to the excluded regions displayed in \reffi{fig:lowMH}. Concerning the
light $\cp$-even Higgs boson, within the yellow region in 
\reffi{fig:lowMH} its mass turns out to be rather low, in the range 
$77 \gev \lsim \Mh \lsim 102\gev$,
i.e.\ significantly below the LEP limit for a SM-like
Higgs~\cite{LEPHiggsSM}. The couplings of the light $\cp$-even Higgs
boson to gauge bosons are heavily suppressed in this region, leading to 
rates for the relevant cross sections that are typically smaller by a
factor of 2--10 than the LEP limits~\cite{LEPHiggsSM}.

%%%%%%%%%%%%%%%%%%%%%%%%% F I G U R E %%%%%%%%%%%%%%%%%%%%%%%%%%%%%%%%%%%%%%%%%
\begin{figure}[htb!]
\begin{center}
\includegraphics[width=0.55\textwidth]{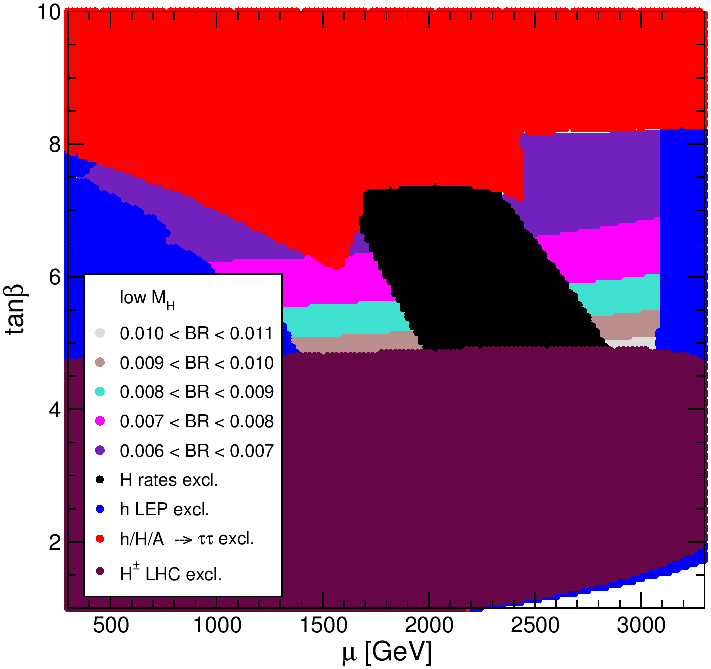}
\caption{
Values of $\br(t \to H^\pm b)$ (denoted as ``BR'') in the $\mu$--$\tb$
plane in the \lowMH\ scenario. The experimenally excluded regions are
indicated as in \reffi{fig:lowMH}.
}
\label{fig:lowMH-tHb}
\end{center}
%\vspace{-2em}
\end{figure}
%%%%%%%%%%%%%%%%%%%%%%%%% F I G U R E %%%%%%%%%%%%%%%%%%%%%%%%%%%%%%%%%%%%%%%%%

While the existing limits from the searches for the MSSM Higgs bosons 
constrain the parameter space of the \lowMH\ scenario, according to our 
assessment based on {\tt HiggsBounds} {\tt 4.0.0} there remains an
interesting parameter region that is unexcluded, as displayed in 
\reffi{fig:lowMH}. The proposed \lowMH\ benchmark scenario
is intended to facilitate a proper experimental analysis that will answer the
question whether scenario giving rise to Higgs phenomenology that is
very different from the SM case is still viable in the MSSM. 
As discussed above, besides the searches for neutral MSSM Higgs bosons
in $\tau^+\tau^-$ final states also charged Higgs searches have a high
sensitivity for probing this scenario. In order to investigate the
prospects for charged Higgs searches in top quark decays in more
detail, we show in \reffi{fig:lowMH-tHb} the predictions 
for $\br(t \to H^\pm b)$ (denoted as ``BR'' in the plot) in the  
unexcluded region of the $\mu$--$\tb$ plane of the \lowMH\  scenario.
One observes that this branching ratio is just below the current
experimental limits \cite{ChargedHiggs}, which are at the level of~$1\%$.

%%%%%%%%%%%%%%%%%%%%%%%%%%%%%%%%%%%%%%%%%%%%%%%%%%%%%%%%%%%%%%%%%%%%%%%%
%%%%%%%%%%%%%%%%%%%%%%%%%%%%%%%%%%%%%%%%%%%%%%%%%%%%%%%%%%%%%%%%%%%%%%%%

%\clearpage
\newpage

\section{Conclusions}

In this paper we have proposed new benchmark scenarios for MSSM Higgs
boson searches at the LHC. The proposed benchmarks are expressed in
terms of low-energy MSSM parameters and are restricted to the
($\cp$-conserving) case of real parameters.
The benchmark scenarios take into account the recent
discovery of a Higgs-like state at $\sim \mass \gev$, i.e.\ over a wide
range of their parameter space they are compatible with both the mass
and the detected production rates of the observed signal. This refers 
to the interpretation of the signal in terms of the light $\cp$-even
Higgs boson of the MSSM, with the exception of the \lowMH\  scenario, 
where the observed signal is interpreted as the heavier $\cp$-even Higgs
boson. For each scenario we have investigated the impact on the
parameter space from the current exclusion bounds from Higgs searches at
LEP, the Tevatron and the LHC (taking both
experimental and theory uncertainties into account). The benchmark
scenarios have been chosen to demonstrate certain features of MSSM
Higgs phenomenology.

The proposed set of benchmarks comprises a slightly updated version of
the well-known \mhmax\ scenario, which can be used to obtain 
conservative lower bounds on $\MA$, $\MHp$
and $\tb$ via the interpretation of the light $\cp$-even Higgs as the
newly observed state at $\sim \mass \gev$ (including theoretical
uncertainties). Furthermore we propose a {\it modified scenario}
(\mhmod), which differs from the \mhmax\ scenario by reducing the mixing
in the stop sector (parametrized by $|\Xt/\msusy|$)
compared to the value that maximizes $\Mh$. Two versions of this
scenario are proposed, one with a positive and one with a negative sign
of $\Xt$. Within (both versions of) the \mhmod\ scenario the light
$\cp$-even Higgs boson can be interpreted as the newly discovered state
within the whole parameter space of the $\MA$--$\tb$ plane that is
unexcluded by limits from Higgs searches at LEP and the LHC, except for
a small region with very small values of $\tb$.
We expect the \mhmod\ scenario to be useful for the future
interpretations of the searches for the heavy MSSM Higgs bosons
$H$, $A$ and $H^\pm$.

As we have discussed in some detail for the \mhmax\ and \mhmod\
scenarios, the searches for the heavy MSSM Higgs bosons $H$ and $A$ 
in the usual channels with SM fermions in the final state are
significantly affected in parameter regions where decays of $H$ and $A$
into supersymmetric particles are possible. In particular, we have
discussed decays into charginos and neutralinos as well as decays into
staus. Furthermore, decays of the heavy $\cp$-even Higgs boson into a
pair of light $\cp$-even Higgs bosons can be important. 
We encourage ATLAS and CMS to enhance the 
sensitivity of their searches
for MSSM Higgs bosons by performing also dedicated searches for Higgs
decays into SUSY particles 
and into a pair of lighter Higgs bosons.

We have also defined
the \lstop\ scenario, which has $\mste \approx 325 \gev$ and
$\mstz \approx 670 \gev$. The stops give a sizable contribution to
the $\si(gg \to h)$ production rate. 
Similarly, we define the \lstau\ scenario, where the light staus can
enhance $\Ga(h \to \ga\ga)$ substantially at high values of $\tb$. 
We have furthermore proposed
the \tauphobic\ scenario, which exhibits potentially sizable
variations of $\Ga(h \to b \bar b)$ and $\Ga(h \to \tau\tau)$
with respect to their SM values.
For the \mhmax, \mhmod\ and \lstop\ scenarios we propose to
investigate several values 
(and in particular both signs) of the parameter $\mu$, which has an
important impact on the bottom Yukawa coupling via the corrections
involving the quantity~$\db$.

Finally, we define the \lowMH\ scenario, which interprets the {\em heavy}
$\cp$-even Higgs boson as the newly discovered state at $\sim \mass \gev$.  
Since this scenario by definition requires a low value of $\MA$, we
keep $\MA$ fixed and instead vary $\mu$ as a free parameter, i.e.\
the $\mu$--$\tb$ parameter space is investigated. In most of the allowed
parameter space the mass of the heavy
$\cp$-even Higgs boson is close to $\mass \gev$, and its production and
decay rates are SM-like. The light $\cp$-even Higgs boson, on the other
hand, has heavily suppressed couplings to gauge bosons and a mass
that is typically below the LEP limit for a SM-like
Higgs. The \lowMH\ scenario is characterized by a particularly rich
phenomenology, since all five MSSM Higgs bosons are
light. Besides the searches for neutral MSSM Higgs bosons
in $\tau^+\tau^-$ final states also charged Higgs boson searches have a high
sensitivity for probing this scenario. This scenario could therefore
serve also as a useful benchmark for (light)
charged Higgs boson searches in the MSSM.

\subsection*{Acknowledgements}

We thank C.~Acereda Ortiz for discussions on the decay rates of 
$H \to hh$ and Y.~Linke for discussions on the \mhmod\ and \lowMH\
scenarios.
We thank P.~Bechtle and T.~Stefaniak for discussions
on {\tt HiggsBounds}.
This work has been supported by the Collaborative Research Center
SFB676
of the DFG, ``Particles, Strings, and the Early Universe''.
The work of S.H.\ was partially supported by CICYT (grant FPA
2010--22163-C02-01) and by the Spanish MICINN's Consolider-Ingenio 2010
Programme under grant MultiDark CSD2009-00064. 
The work of O.S.\ is supported by the Swedish Research Council (VR)
through the Oskar Klein Centre.
Fermilab is operated by Fermi Research Alliance, LLC under Contract No.\
DE-AC02-07CH11359 with the U.S.\ Department of Energy. Work at ANL is
supported in part by the U.S.\ Department of Energy under Contract No.\
DE-AC02-06CH11357.

%%%%%%%%%%%%%%%%%%%%%%%%%%%%%%%%%%%%%%%%%%%%%%%%%%%%%%%%%%%%%%%%%%%%%%%%
%%%%%%%%%%%%%%%%%%%%%%%%%%%%%%%%%%%%%%%%%%%%%%%%%%%%%%%%%%%%%%%%%%%%%%%%

\newpage
%\appendix
\section*{Appendix: Summary of parameter values}

\begin{table}[b!]
\centering
\includegraphics[width=1.12\columnwidth,angle=90]{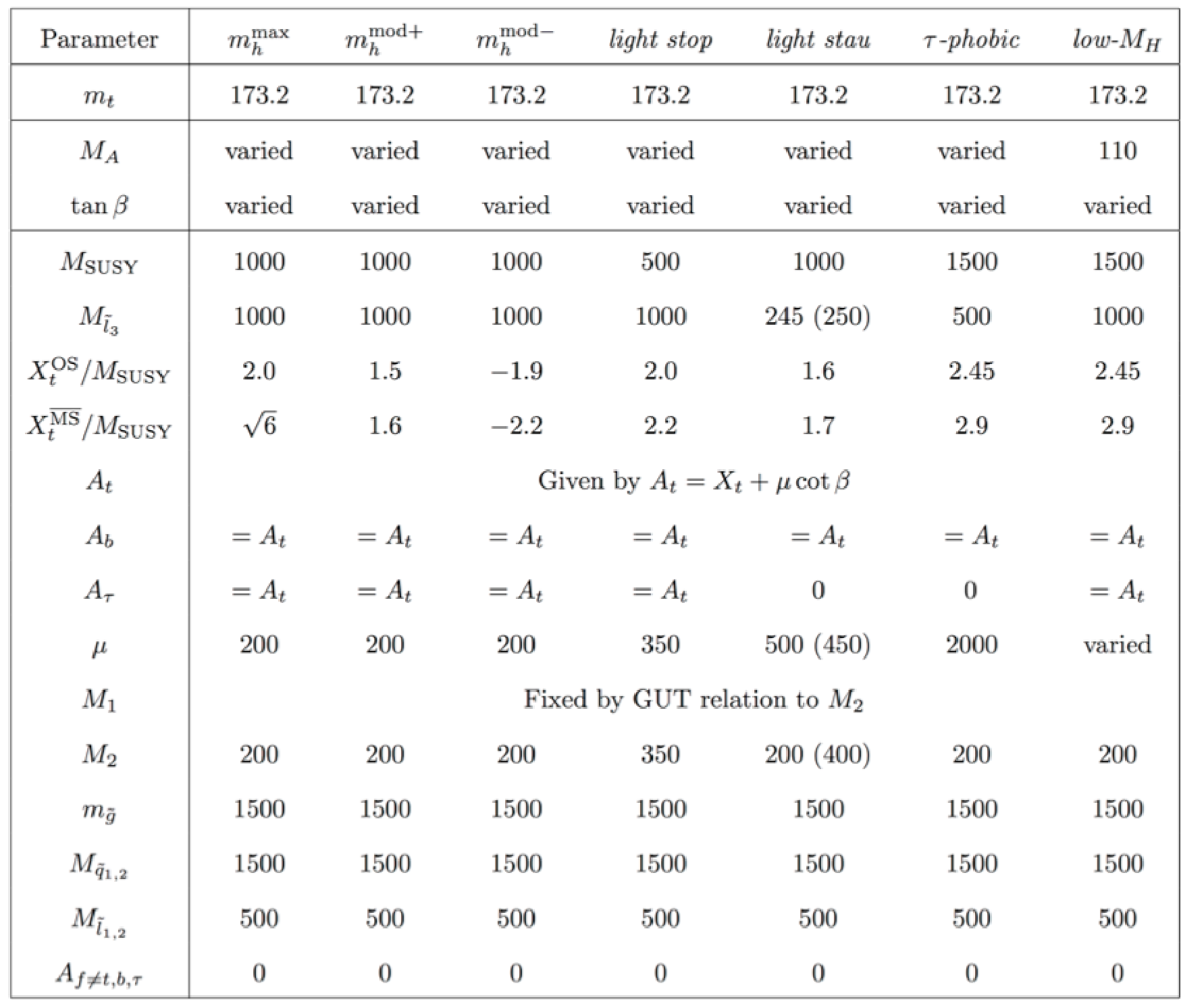}
\caption{Summary of parameter values for the proposed
benchmark scenarios,
  given in the on-shell (OS) scheme unless otherwise noted. Numbers in
  parentheses refer to calculations with $\Delta_\tau$ effects included
  in the stau mass evaluation
  (see the description of the \lstau\ scenario for
  details). Dimensionful quantities are given in GeV.}
\end{table}

%%%%%%%%%%%%%%%%%%%%%%%%%%%%%%%%%%%%%%%%%%%%%%%%%%%%%%%%%%%%%%%%%%%%%%%%
%%%%%%%%%%%%%%%%%%%%%%%%%%%%%%%%%%%%%%%%%%%%%%%%%%%%%%%%%%%%%%%%%%%%%%%%

%%%%%%%%%%%%%%%%%%%%%%%%%%%%%%%%%%%%%%%%%%%%%%%%%%%%%%%%%%%%%%%%%%%%%%%%
%%%%%%%%%%%%%%%%%%%%%%%%%%%%%%%%%%%%%%%%%%%%%%%%%%%%%%%%%%%%%%%%%%%%%%%%

\newpage
\bibliographystyle{plain}

%%%%%%%%%%%%%%%%%%%%%%%%%%%%%%%%%%%%%%%%%%%%%%%%%%%%%%%%%%%%%%%%%%%%%%%%
%%%%%%%%%%%%%%%%%%%%%%%%%%%%%%%%%%%%%%%%%%%%%%%%%%%%%%%%%%%%%%%%%%%%%%%%

\end{document}